\documentclass[12pt,english]{article}

\usepackage{amsmath,amssymb }
\usepackage{fancybox}
\usepackage{graphicx,psfrag,epsf}
\usepackage{enumerate}
\usepackage{graphicx,psfrag}
\usepackage{multirow}
\usepackage{epsfig}
\usepackage{subfigure}
\usepackage{theorem}
\usepackage{natbib,psfrag}
\usepackage[usenames,dvipsnames]{color}
\usepackage{todonotes}
\usepackage{xcolor}
\usepackage{algorithm2e}
\usepackage{tabularx}
\usepackage{xr}
\usepackage{changepage}
\usepackage{geometry}
\usepackage{subfig}
\usepackage[font=small,skip=2pt]{caption}
\newcommand{\pdf}{0}
\if1\pdf
\usepackage[pdftex]{graphicx}
\else
\usepackage{graphicx}
\fi

\newcommand{\blind}{1}

\def\mb#1{\setbox0=\hbox{$#1$}
  \kern-.025em\copy0\kern-\wd0
  \kern.05em\copy0\kern-\wd0
  \kern-.025em\raise.0em\box0}

\makeatletter
\newcommand*\dashline{\rotatebox[origin=c]{90}{$\dabar@\dabar@\dabar@$}}
\makeatother

\newcommand{\pr}{\text{Pr}}

\newcommand{\diag}{\text{diag}}

\begin{document}

\baselineskip 1.8em
\parskip 1em

%
%

\def\spacingset#1{\renewcommand{\baselinestretch}%
	{#1}\small\normalsize} \spacingset{1}

\if1\blind
{
	\title{\bf Adaptive Bayesian Spectral Analysis of High-dimensional Nonstationary Time Series}
	\author{Zeda Li,  Ori Rosen, Fabio Ferrarelli and Robert T. Krafty
		\footnote{Zeda Li is Assistant Professor, Paul H. Chook Department of Informartion System and Statistics, Baruch College, The City University of New York (zeda.li@baruch.cuny.edu), Ori Rosen is Professor, Department of Mathematical Science, University of Texas at El Paso (orosen@utep.edu), Fabio Ferrarelli is Assistant Professor, Department of Psychiatry, University of Pittsburgh (ferrarellif@upmc.edu), and Robert T. Krafty is Associate Professor, Department of Biostatistics, University of Pittsburgh (rkrafty@pitt.edu).}}
	\maketitle
} \fi

\if0\blind
{
	\bigskip
	\bigskip
	\bigskip
	\begin{center}
		{\LARGE\bf Adaptive Bayesian Spectral Analysis of High-dimensional Nonstationary Time Series}
	\end{center}
	\medskip
} \fi

\newpage

\setlength{\baselineskip}{22pt}  

\begin{center}
\section*{Abstract}
\end{center}

This article introduces a nonparametric approach to spectral analysis of a high-dimensional multivariate nonstationary time series.  The procedure is based on a novel frequency-domain factor model that provides a flexible yet parsimonious representation of spectral matrices from a large number of simultaneously observed time series. Real and imaginary parts of the factor loading matrices are modeled independently using a prior that is formulated from the tensor product of penalized splines and multiplicative gamma process shrinkage priors, allowing for infinitely many factors with loadings increasingly shrunk towards zero as the column index increases. Formulated in a fully Bayesian framework, the time series is adaptively partitioned into approximately stationary segments, where both the number and location of partition points are assumed unknown. Stochastic approximation Monte Carlo (SAMC) techniques are used to accommodate the unknown number of segments, and a conditional Whittle likelihood-based Gibbs sampler is developed for efficient sampling within segments.  By averaging over the distribution of partitions, the proposed method can approximate both abrupt and slowly varying changes in spectral matrices.
Performance of the proposed model is evaluated by extensive simulations and demonstrated through the analysis of high-density electroencephalography.

\noindent%
KEY WORDS: Factor Analysis; High-dimensional Time Series; Locally Stationary Process; Multiplicative Gamma Process; Penalized Splines; Spectral Analysis; Stochastic Approximation Monte Carlo.

\bigskip

\section{Introduction}\label{subsec:intro}

Technological advances have facilitated an explosion in the number of studies that simultaneously record a large number of processes over time.  In many applications, important scientific information is contained in the variability attributable to oscillations at different frequencies, which can be quantified through the power spectrum.
Simultaneous analyses of such data, which take into account all cross-time series dependencies as well as within-series frequency patterns, provide a  comprehensive understanding of the underlying process. However, such analyses are challenging when the data are high-dimensional and nonstationary, since unwieldy sizes of spectral matrices evolving over time present methodological and computational obstacles.

\begin{figure}[ht]
	\centering
	\includegraphics[height=2.5in]{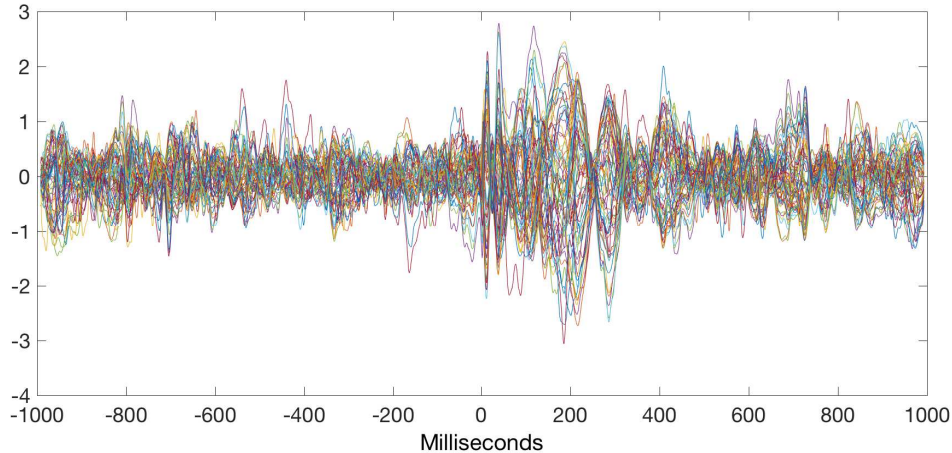}
	\caption{64-channel TMS-evoked hdEEG from a first-episode psychosis patient.}
	\label{tseeg}
\end{figure}
Our motivating application comes from the analysis of 64-channel high-density electroencephalography (hdEEG) during transcranial magnetic stimulation (TMS) in  a patient who was hospitalized during a first-break psychotic episode (Figure \ref{tseeg}).   TMS is a noninvasive procedure that uses magnetic fields to excite brain cells.  EEG is used to measure electrophysiological activity simultaneously across multiple regions, or channels, of the brain.  Traditional EEG measures activity at 2-16 different channels.   More recently, hdEEG, which can measure activity from up to 256 channels, has been utilized to obtain a more comprehensive view of brain activity.
The power spectrum of hdEEG during TMS and its evolution provide dynamic, high-resolution neurobiological information with regards to neurological mechanisms, disruptions of which have been observed in patients with a clinical diagnosis of schizophrenia \citep{kaskie2018}.   The goal of our analysis is to investigate hdEEG during TMS in our patient who is experiencing a first-break psychosis, a potential precursor to a future diagnosis of schizophrenia, which could serve as a potential pre-clinical biomarker \citep{ferrarelli2018}.

Historically, a rolling-window procedure \citep{priestley1981} is used for the spectral analysis of nonstationary time series whose second-order structure evolves over time. This procedure partitions the time series into prespecified overlapping time blocks and then estimates the power spectrum at each of these blocks by smoothing the periodogram, which is a noisy estimate of the power spectrum, across frequencies using tools such as local averaging \citep{shumway2011}.
More recently, a variety of approaches to the spectral analysis of multivariate nonstationary time series have been proposed. These methods can be roughly grouped into three categories:
methods that assume, in a manner similar to the rolling window estimator,  power spectra evolve smoothly over time \citep{dahlhaus2000, guo2006, sanderson2010, park2014},  methods that are based on data-driven piecewise stationary approximations  \citep{ombao2005, davis2006}, and adaptive Bayesian methods that can approximate both abrupt and slowly varying dynamics \citep{zhang2016, li2018}.
However, these methods focus on the analysis of low-dimensional collections of time series, becoming theoretically unjustified or computationally infeasible when a large, or even moderate, number of simultaneous series are observed. The primary contribution of this article is introducing an adaptive method for the spectral analysis of high-dimensional time series that can capture both abrupt and slowly varying dynamics efficiently while nonparmetricaly modeling spectral matrices.

Various approaches have been developed for estimating the spectral matrix of a high-dimensional stationary time series. These include shrinkage estimators  \citep{bohm2009, fiecas2010, fiecas2014,luftman2016} and thresholding estimators \citep{sun2018,fiecas2018}. 
 Unfortunately, these methods are not readily extendable to the spectral analysis of nonstationary time series. A common approach to analyzing  high-dimensional time series is to use factor models, as they not only induce a parsimonious and interpretable structure among the time series, but can also overcome the curse of dimensionality when estimating high-dimensional covariance or spectral matrices, see \cite{ensor2013} for a review. Although factor models have been used extensively for time-domain time series analysis, mostly in the econometrics literature, existing factor models for frequency-domain analysis of time series are relatively few. These methods are primarily framed in the stationary setting \citep{stoffer1999, prado2014}, or are based on time-domain parametric formulations \citep{west1999, prado2001}.  In addition to these time series factor models, factor models for the analysis of functional time series can also be adapted for the analysis of multivariate power spectra \citep{kowal2017}. However, the direct application of such functional data-based procedures for spectral analysis can only estimate predefined functions, such as univariate log-spectra and squared coherences, but not the entire spectral matrix.



In this article, we present an adaptive nonparametric method for the spectral analysis of nonstationary high-dimensional time series.  Our approach is based on a novel frequency- domain locally stationary factor model and scalable Markov chain Monte Carlo (MCMC) techniques. The factor model is general in the sense that it allows for both real- and complex- valued spectral matrices, which means that individual time series can fluctuate simultaneously or propagate in a lagged fashion. Real and imaginary parts of the local factor loading matrices are modeled independently  using a novel prior formulated through the tensor product of penalized spline priors, which induce smooth structure across frequency, and  multiplicative gamma process shrinkage priors \citep{Bhattacharya2011}, which allow infinitely many factors with the loadings increasingly shrunk towards zero as the column index increases and mitigate the need for estimating the number of factors.  The approach adaptively divides a high-dimensional time series into an unknown random number of approximately stationary segments of variable lengths through stochastic approximation Monte Carlo (SAMC). By approximating the likelihood function via products of local Whittle likelihoods, a scalable and efficient Gibbs sampling algorithm is developed. As in other adaptive Bayesian methods \citep{rosen2012, zhang2016, li2018}, by averaging estimates across the distribution of partitions, the method not only produces estimates that can capture abrupt changes, but also effectively approximates slowly varying processes.

The article is organized as follows. Section 2 presents the proposed frequency-domain factor model for high-dimensional nonstationary time series. Section 3 describes the prior distributions for the model parameters and an outline of the sampling scheme. Section 4 presents the results of extensive simulation studies. The proposed method is applied to real data in Section 5.  Section 6 concludes with some discussion and future work. A detailed description of the sampling scheme is given in the supplementary material.

\section{The Model}\label{sec:model}

\subsection{The Locally Stationary Factor Model}\label{subsec:factor}

This article considers $P$-dimensional vector-valued time series $\mb X_t = (X_{1t}, \cdots, X_{Pt})'$ that can be represented through a novel locally stationary factor model. The model possesses a Cram\'{e}r representation with time-varying factor loadings, which decomposes $\mb X_t$ into information accounted for by a set of $Q$ common processes, or factors, plus an idiosyncratic component. Formally, we model an $\mathbb{R}^P$-valued time series of length $T$, $\{ \mb X_t: t=1, \cdots, T \}$, as
\begin{equation}\label{eq:model}
\mb X_t =\int_{0}^{1} \Lambda(t/T, \omega) \exp{(2 \pi i \omega t)}  d \mb Z(\omega) + \mb \epsilon_t,
\end{equation}
where $\mb Z(\omega)$ is a $Q$-dimensional mean-zero orthogonal incremental process with independent and Hermitian latent factors, such that  $E\left\{  d \mb Z(\omega) d \mb Z^*(\zeta) \right \}$ is the identity matrix if $\omega = \zeta$ and zero otherwise, $\mb \epsilon_t$ is a  $P$-dimensional independent white noise, and $\Lambda(u, \omega)$ is a $P \times Q$ time-varying loading matrix that is a function of scaled time $u \in [0,1]$ and frequency $\omega \in \mathbb{R}$. The complex-valued loading matrix $\Lambda(u, \omega)$ is periodic and Hermitian as a function of frequency such that $\Lambda(u, \omega) = \Lambda(u, \omega + 1)$ and $\Lambda(u, \omega) = \overline{\Lambda(u, -\omega)}$, where $\overline{\Lambda}$ is the conjugate of $\Lambda$.
The time-varying power spectrum defined through the locally stationary factor Cram\'{e}r representation in \eqref{eq:model} is given by
\begin{equation*}
\label{spect}
f(u, \omega) = \Lambda(u, \omega)\Lambda(u, \omega)^* + \Sigma_{\epsilon}, \quad u \in [0,1],~ \omega \in \mathbb{R},
\end{equation*}
where $\Lambda^*$ is the conjugate transpose of $\Lambda$, and $\Sigma_{\epsilon}$ is the $P \times P$ diagonal covariance matrix of $\mb \epsilon_t$. Hence, the time-varying spectrum $f(u,\omega)$ is a complex-valued positive definite $P \times P$ Hermitian matrix.  We assume that given $u$,  each component of $f(u, \cdot)$  possesses a square-integrable first derivative as a function of frequency; given $\omega$, each component of $f(\cdot, \omega)$ is continuous as a function of scaled time at all but a possibly finite number of points.

Several aspects of the proposed locally stationary factor model should be noted.  First,
a time series defined by the locally stationary factor model is locally stationary in the sense of \cite{li2018}, which differs slightly from the definitions of local stationarity used by \cite{dahlhaus2000} and \cite{guo2006}.  The model of \cite{dahlhaus2000} assumes a series of transfer functions/loadings indexed by the length of the time series that converges to a fixed function as the length increases.  This structure is adopted primarily to allow for the fitting of popular parametric models such as time-varying moving average models.  Since we are considering nonparametric estimation, in a manner similar to that in \cite{guo2006} and \cite{li2018}, we directly use the limiting function.  Further, similar to the method of \cite{li2018}, the proposed model allows for a finite number of discontinuities of the spectrum as a function of time.  This is more flexible than the models of \cite{dahlhaus2000} and \cite{guo2006}, which require continuity as a function of time.  Second,  through the introduction of the tensor product penalized spline and multiplicative gamma process shrinkage prior (see details in Section \ref{subsec:local}), the proposed model allows for an infinite number of factors with sufficiently decaying loadings.  This enables any locally stationary process in the sense of \cite{guo2006} or \cite{li2018} to be represented as a locally stationary factor model.   The use of this prior mitigates the sensitivity to the number of factors.  Nevertheless, the finite factor representation is used to overcome the curse of dimensionality in estimating the power spectrum, where the number of factors $Q$ is typically smaller than $P$, inducing a reduced-rank characterization of the spectral matrix. Lastly, a common concern about any factor analysis is that the loadings are not identifiable without appropriate constraints, such as lower triangular structure for the loading matrix \citep{geweke1996, carvalho2008}.  It should be emphasized that our goal is not to identify or interpret the factors themselves, but to estimate time-varying power spectral matrices.  As noted by \cite{Bhattacharya2011} in the context of covariance matrices, the shrinkage provided by the multiplicative gamma process prior enables valid estimation, prediction and inference on power spectrum despite the lack of identifiability of the loadings.

\subsection{Piecewise Stationary Approximation}

A locally stationary time series can be accurately approximated as a piecewise stationary process \citep{adak1998, guo2006}, and our procedure involves a piecewise stationary approximation to the locally stationary factor model. For a time series of length $T$ $\left\{ \mb X_t : t =1, \dots, T \right\}$, a collection of partition points of the time series into $L$ segments is denoted by $\mb \xi = \left(\xi_{0}, \dots, \xi_{L}\right)'$  where $\xi_{0} = 0$ and $\xi_{L} = T$ such that $\mb X_t$ is approximately stationary within the segments $\left\{t : \, \xi_{\ell-1} < t \le \xi_{\ell}\right\}$ for $\ell=1, \dots, L$. Conditional on $L$ and $\mb \xi$,
\begin{equation*}
\mb X_t \approx \sum_{\ell=1}^{L} \int_{0}^{1} \Lambda_{\ell}(\omega) \exp{(2 \pi i \omega t)} d \mb Z(\omega) + \mb \epsilon_t,
\end{equation*}
where for $\ell = 1, \cdots, L$, $\Lambda_{\ell}(\omega)=\Lambda(u_{\ell}, \omega) \delta_{\ell}(t)$, $\delta_{\ell}(t)$ is an indicator function such that $\delta_{\ell}(t) = 1$ if $t \in (\xi_{\ell-1}, \xi_{\ell}]$ and zero otherwise, and $u_{\ell} = \left( \xi_{\ell} + \xi_{\ell-1}\right)/2$ is the midpoint of the $\ell$th segment.
Within the $\ell$th segment, the time series is approximately second-order stationary with local power spectrum $f(u_{\ell}, \omega)= \Lambda_{\ell}(\omega)\Lambda_{\ell}(\omega)^* + \Sigma_{\epsilon}$, where  $\Sigma_{\epsilon}$ is a diagonal covariance matrix of $\mb \epsilon_t$. It should be noted that the number of segments $L$ and partition $\mb \xi$ are random variables whose prior distributions are given in Section 3.2. Estimates and inferences will be obtained by averaging over the posterior distribution of the partitions and the number of segments.

Conditional on approximately stationary segments, define the local discrete Fourier transform (DFT) at frequency $k$ within segment $\ell$ as
\begin{equation*}
\mb Y_{k \ell} =  T_{\ell}^{-1/2} \sum_{t = \xi_{\ell-1} + 1}^{\xi_{\ell}} \mb X_{t} \exp(-2 \pi i  \omega_{k \ell} t), \quad k=1, \dots, K_{\ell}, \, \ell=1, \dots, L,
\end{equation*}
where $T_{\ell}$ is the number of time points in the $\ell$th segment, $\omega_{k \ell} = k/T_{\ell}$, $k=1, \cdots K_{\ell}$ are the Fourier frequencies, and  $K_{\ell} = \lfloor(T_{\ell}-1)/2\rfloor$, which is the greatest integer that is less than or equal to $(T_{\ell}-1)/2$. Under some regularity conditions \citep[Theorem 4.4.1]{brillinger2001}, the $\mb Y_{k\ell}$ are approximately independent multivariate complex Gaussian random vectors with mean $\mb 0$ and covariance $f(u, \omega_{k \ell})$, denoted by $\mb Y_{k\ell} \overset{\text{app}}{\sim} CN[\mb 0, f(u, \omega_{k \ell})]$. The sampling scheme follows from the distribution of the local DFTs conditional on the latent factors. In particular, we let $\mb D_{k \ell} = \int_0^{\omega_{k \ell}} \mb Z(\omega) d\omega$, $\mb E_{k\ell} = T_{\ell}^{-1/2} \sum_{t=\xi_{\ell-1}}^{\xi_{\ell}} \mb \epsilon_t \exp(-2\pi i \omega_{k \ell}t)$, and $\Lambda_{k\ell} = \Lambda(u_{\ell}, \omega_{k \ell})$. It then follows that \citep{brillinger2001}
\begin{equation*}
\mb Y_{k \ell} \approx \Lambda_{k \ell} \mb D_{k \ell} + \mb E_{k \ell},
\end{equation*}
where $\mb D_{k \ell}$ is approximately independent $CN(\mb 0, I_Q)$, $I_Q$ is a $Q \times Q$ identity matrix, and $\mb E_{k \ell} \sim CN(\mb 0, \Sigma_{\epsilon})$. This leads to the conditional Whittle likelihood
\begin{equation*}\label{likeihood1}
{\cal L}( \mb Y \mid \mb \Lambda, \mb D, \Sigma_{\epsilon}, L) \approx  \prod_{\ell=1}^{L} \prod_{k=1}^{K_\ell} \prod_{p=1}^P \left\{  \sigma_{\epsilon,  p}^{-1} \exp \left (  -\sigma_{\epsilon, p}^{-2} \left |  Y_{k \ell p} - \Lambda_{k\ell}^{(p)} \mb D_{k \ell}  \right |^2 \right )  \right\},
\end{equation*}
where $\mb Y$ represents all local discrete Fourier transforms, $\mb \Lambda$ and $\mb D$ represent the collections of loadings and factors at all segments and Fourier frequencies, $\sigma_{\epsilon,p}^2$ is the $p$th diagonal element of $\Sigma_{\epsilon}$, $Y_{k\ell p}$ is the $p$th element of $\mb Y_{k \ell}$ and $\Lambda_{k \ell}^{(p)}$ is the $p$th row of $\Lambda_{k \ell}$.


\section{Adaptive Bayesian Spectral Analysis}\label{sec:analysis}

In this section, we introduce an adaptive Bayesian approach that extends the approaches of \cite{rosen2012}, \cite{zhang2016} and \cite{li2018} to spectral analysis of high-dimensional time series. Under this approach, a time series is adaptively partitioned into a random number of approximately stationary segments, local spectra are estimated within each such segment, and time-varying power spectrum is obtained by averaging local estimates over the distribution of partitions.  First,  in Section \ref{subsec:local}, we  present the estimation procedure for a high-dimensional {\em stationary} time series.
Then, in Section \ref{subsec:adaptive}, we introduce our proposed SAMC-based adaptive Bayesian sampling scheme for temporal partitioning of a {\em nonstationary} time series.

\subsection{Spectral Estimation for Stationary Time Series}\label{subsec:local}

To aid the presentation, in this section, 
we focus on estimating the spectral matrix, $f(\omega)$, of a stationary time series.
Under the factor formulation, $f(\omega)$ is expressed as a sum of a nonnegative definite matrix $\Lambda(\omega)\Lambda(\omega)^*$ induced by the loadings, and a positive-definite diagonal matrix $\Sigma_{\epsilon}$.


Given the factor model, the conditional Whittle likelihood for a stationary multivariate time series is given by
\begin{equation*}\label{likeihood2}
{\cal L}( \mb Y \mid  \Lambda, \mb D, \Sigma_{\epsilon}) \approx   \prod_{k=1}^{K} \prod_{p=1}^P \left\{  \sigma_{\epsilon,  p}^{-1} \exp \left (  -\sigma_{\epsilon,  p}^{-2} \left |  Y_{k p} - \Lambda_{k}^{(p)} \mb D_{k}  \right |^2 \right )  \right\},
\end{equation*}
where $\mb Y$ is the discrete Fourier transforms of the time series, $\mb \Lambda$ and $\mb D$ denote the collections of loadings and factors, respectively, $Y_{k p}$ is the $p$th element of $\mb Y_{k}$ and $\Lambda_{k}^{(p)}$ is the $p$th row of $\Lambda_{k}$. To complete our model, we place prior distributions on $\mb \Lambda$ and $\sigma_{\epsilon,p}$.

Prior distributions on the error variances are placed by assuming that the $\sigma_{\epsilon,p}$, for $p=1, \cdots, P$, are independent Half-$t(\nu, G_{\epsilon})$ with density $p(\sigma_{\epsilon,p}) \propto [1 + (\sigma_{\epsilon,p}/G_{\epsilon})^2/\nu]^{-(\nu+1)/2}$, $\sigma_{\epsilon,p}>0$, where the hyperparameters $\nu$ and $G_{\epsilon}$ are known constants \citep{gelman2006}. In practice, this distribution can be represented by a scale mixture of inverse gamma distributions \citep{wand2012}: $(\sigma|g_{\epsilon,p}) \sim \text{IG}(\nu/2, \nu/g_{\epsilon,p})$, $g_{\epsilon,p} \sim \text{IG}(1/2, 1/G_{\epsilon,p}^2)$, where $\text{IG}(a,b)$ denotes an inverse Gamma distribution with density $p(x) \propto x^{-(a+1)}\exp(-b/x)$, $x>0$.

A novel prior is placed on the $P \times Q$ complex-valued loading matrix $\Lambda(\omega)$, such that the real and imaginary parts of its entries are modeled independently through tensor products of Bayesian penalized splines \citep{krafty2018, li2018} and multiplicative gamma process shrinkage priors \citep{Bhattacharya2011}. In particular, we use the first $S$ Demmler-Reinsch basis functions
\begin{eqnarray}
	\label{one}
\Re\{\Lambda_{p q}(\omega)\}\ &=& \alpha_{p q 0} + \sum_{s=1}^{S-1} \alpha_{p q s} \sqrt{2} \cos(2\pi s \omega), \\
	\label{two}
\Im\{\Lambda_{p q}(\omega)\}\ &=& \sum_{s=1}^{S} \beta_{p q s} \sqrt{2} \sin(2\pi s \omega).
\end{eqnarray}
In this formulation, the real and imaginary parts of the loading matrix are modeled by periodic even and odd linear splines, respectively, which has been found to improve performance compared to a model that only accounts for periodic patterns but does not restrict the functions to be odd or even \citep{krafty2013}.  A Bayesian penalized spline can be formulated by placing independent $N\left[0, \left(2 \pi s\right)^{-1} \tau^2\right]$ priors on the coefficients conditional on a smoothing parameter $\tau^2$.  This prior induces smoothness as a function of frequency by regularizing integrated squared first derivatives.  However, we also desire the loadings to decay for large $q$, so that the majority of information is captured by the first several factors.  To achieve this, we introduce a prior that is a tensor product of this penalized spline prior with a gamma process shrinkage prior.  Formally, we define the prior distributions as:
\begin{enumerate}
\item $\alpha_{p q 0} \sim N\left (0,\psi_{q, (re)}^{-1}\right)$, $\alpha_{p q s} \sim N\left (  0, (2 \pi  s)^{-1}  \tau_{ p q, (re)}^2 \psi_{ q, (re)}^{-1} \right )$ for $s=1, \cdots, S-1$.
\item $\beta_{p q s} \sim N\left (  0, (2 \pi  s)^{-1}  \tau_{p q, (im)}^2 \psi_{q, (im)}^{-1} \right )$ for $s=1, \cdots, S$.
\item $\tau_{p q s, (re)}, \tau_{p q s, (im)}\sim$ Half-$t(\nu, G_{\tau}^2)$ are the square roots of the smoothing parameters of the real and imaginary parts, respectively, which control the roughness as a function of frequency.
\item $\psi_{q, (\cdot)}$, $q=1, \cdots, Q$, and $(\cdot)$=$ (re)~\text{or}~(im)$, are shrinkage parameters that control the decay of the columns of the loading matrix, where $\psi_{ q, (\cdot)} = \prod_{h=1}^q \phi_{h,(\cdot)}$, with independent priors $\phi_{1, (\cdot)} \sim \text{Ga}(a_1,1)$, $\phi_{ q, (\cdot)} \sim \text{Ga}(a_2,1)$ for $q \ge 2$, where $a_1$ and $a_2$ are fixed constants.
\end{enumerate}

This formulation has three favorable properties. First, the choice of $S$ provides a compromise between loss of accuracy relative to the full rank Bayesian smoothing spline  ($S=K$) and computational feasibility. Suggested by \cite{krafty2018} and \cite{li2018}, we select $S=10$ in our simulation studies and data analysis for considerable computational savings without sacrificing model fit.
Second, the shrinkage parameters $\psi_{q, (\cdot)}$ are stochastically increasing when  $a_2 >1$, which places more shrinkage toward zero as the column index increases.
This formulation allows an infinite number of factors with sufficiently decaying loadings, which not only provides a good approximation to the spectral representation of multivariate stationary time series \citep{brillinger2001}, but also reduces the sensitivity to the number of factors. Third, this formulation allows for sampling from the posterior distribution through an efficient Gibbs sampler which scales well to high-dimensional time series.  The sampling scheme can also be performed in parallel to further increase computational speed. Details of the sampling scheme are provided in the supplementary material.

\subsection{Adaptive Spectral Estimation for Nonstationary Time Series}\label{subsec:adaptive}

As described in Section 2, for spectral analysis of nonstationary time series, we adaptively estimate the unknown number and locations of the partition points. Conditional on the partitions, local estimation is performed as in Section \ref{subsec:local}.

\subsubsection{Priors}

We first specify prior distributions for the number and locations of the partitions. Following \cite{rosen2012}, a discrete uniform prior is placed on the number of partitions, such that $L \sim \mathcal{U}(1,L_{\max})$,  where $L_{\max}$ is a fixed large integer that represents the maximum possible number of possible segments.   In general, $L_{\max}$ is set to be large enough to capture all possible approximately stationary segments, but if after running the procedure, we find that the conditional probability of $L_{\max}$ is not approximately zero,
we increase $L_{\max}$. To ensure that the large-sample local Whittle likelihood approximation holds, we choose a minimum number of time points per segment, denoted by $t_{\min}$.
Given the number of segments $L$, equal weights on all possible locations of a partition point conditional on previous partition points are placed. Specifically, the prior for the partition $\mb \xi$ is
\begin{equation*}
\pr(\mb \xi \mid L) = \prod_{\ell=1}^{L-1} \pr(\xi_{\ell} \mid \xi_{\ell-1}, \ell),
\end{equation*}
where $\pr(\xi_{\ell} \mid \xi_{\ell-1}, \ell) = 1/a_{\ell}$, and $a_{\ell}=T-\xi_{\ell-1}-(L-\ell+1)t_{\text{min}}+1$ is the number of possible locations for the $\ell$th partition point.

\subsubsection{Sampling Scheme}\label{subsec:sampling}

We develop a stochastic approximation Monte Carlo \citep[SAMC]{liang2007} sampling scheme for automatic and adaptive estimation of the time-varying spectrum. Although reversible jump Markov chain Monte Carlo (RJMCMC) has been shown to be effective for adaptive spectral estimation of low-dimensional multivariate nonstationary time series \citep{li2018}, the self-adjusting property of SAMC is more suitable for high-dimensional nonstationary time series.  We briefly describe the sampling scheme in this section and provide technical details in the supplementary material.

To allow a more compact presentation, we define $\Xi$ as the collection of all parameters, including coefficients of basis functions, smoothing parameters, and shrinkage parameters. The SAMC sampling scheme at iteration $j$ simulates parameters from the following distribution
\begin{equation*}
p_{\vartheta_{j, \ell}}(L, \Xi) \propto \sum_{\ell=1}^{L_{max}}  \frac{p(L, \Xi)}{e^{\vartheta_{j, \ell}}} \delta(L=\ell),
\end{equation*}
where $\vartheta_{j,\ell} \in \Theta$, for all $j$, and $\Theta$ is a compact set. We consider moving the current value of the chain $(L^{j}, \Xi^j_{L^{j}})$ to a proposed value $(L^{j+1}, \Xi^{j+1}_{L^{j+1}})$. Each iteration consists of two types of moves, within-model and between-model moves, which we outline below.
\begin{enumerate}
	\item Between-model moves: propose $L^{j+1}$ by letting $L^{j+1} = L^j + 1$ or $L^j - 1$.
	\begin{itemize}
		\item If $L^{j+1} = L^j + 1$, a birth move is proposed. A new partition point is drawn by first randomly selecting a current segment to split, and then randomly proposing a new partition point within that segment. Conditional on $L^{j+1}$ and the new partition, new smoothing and shrinkage parameters are proposed. Then, factors and  coefficients of the real and imaginary parts of the basis functions are updated by using \eqref{D}, \eqref{RE}, and \eqref{IM} in the supplementary materials, respectively.
		\item Or, if $L^{j+1} = L^j - 1$, a death move is proposed. A partition point is randomly selected to be deleted and a new segment is formed by combing the two adjacent segments separated by this partition point. Then, the current smoothing and shrinkage parameters of these two segments are used to form new smoothing and shrinkage parameters in the newly combined segment. Finally, factors and coefficients of the real and imaginary parts  of the basis functions are updated by \eqref{D}, \eqref{RE}, and \eqref{IM} in the supplementary materials, respectively.
		\item Let $q(L^{j+1}, \Xi^{j+1}_{L^{j+1}} \mid L^j, \Xi^j_{L^j})$ be the proposal density, the proposed move is then accepted with probability
		\begin{equation*}
		A = \min \left \{1,\frac{e^{\vartheta_{j,L^j}}}{e^{\vartheta_{{j+1},L^{j+1}}}} \frac{\pi(L^{j+1}, \Xi_{L^{j+1}}^{j+1} \mid \mb x) \times q (L^j, \Xi_{L^j}^j \mid L^{j+1},  \Xi^{j+1}_{L^{j+1}})}{\pi(L^j, \Xi_{L^j}^j \mid \mb x) \times q (L^{j+1}, \Xi_{L^{j+1}}^{j+1} \mid L^j, \Xi_{L^j}^j)}   \right \},
		\end{equation*}
		where $\mb x$ is the time series, $\pi(\cdot)$ is the posterior distribution,
		and $e^{\vartheta_{j,L^j}}$ and $e^{\vartheta_{{j+1},L^{j+1}}}$ are self-adjusting factors.
		\item Set $\vartheta^* = \vartheta_{j} + \gamma_{j+1} (\mb e_{j+1} - p_0)$, where $p_0 = 1/L_{\max}$ is a predefined probability,
		\{$\gamma_{j}$\} is a positive sequence such that
		$\gamma_{j} = j_0/\max(j_0,j)$, $j_0$ is a predefined constant, $\mb e_{j+1} = (e_{j+1,1}, \cdots, e_{j+1,L_{\max}})'$ and $e_{j+1, \ell} =1$ if $L^{j+1} = \ell$, and zero otherwise. If $\vartheta^* \in \Theta$, then we set $\vartheta_{j+1}= \vartheta^*$; otherwise, set $\vartheta = \vartheta^* + \mb c^*$, where $\mb c^*$ is chosen such that $\vartheta^*+ \mb c^* \in \Theta$.
	\end{itemize}	

	\item Within-model moves: involve no change in the number of segments, i.e. $L^{j+1} = L^j$. A partition point $\xi_{\ell}$ is randomly selected to be relocated. Then, the factors and basis function coefficients are drawn. These two moves are jointly accepted or rejected in a Metropolis-Hasting step. Lastly, smoothing parameters, shrinkage parameters, and error variances are updated via Gibbs sampling steps.
\end{enumerate}

\section{Simulation Studies}\label{sec:example}

In this section, we evaluate the proposed method through simulated datasets. In Section \ref{subsec:spect}, we investigate performance in estimating time-varying spectra of a piecewise stationary time series with one partition as well as of a slowly varying process. In Section \ref{subsec:partitions}, we investigate performance in accurately identifying partition points, both for a stationary time series, as well as for a piecewise stationary time series with multiple partition points.


\subsection{Time-Varying Power Spectrum Estimation}\label{subsec:spect}

Two processes are considered: piecewise stationary and slowly varying. For each process, we fix the number of observations at $T=2048$, and consider two different dimensions: $P=24$ and $P=48$. For each simulated time series, the time-varying power spectrum is estimated by two methods. First, the proposed method with a total of 10000 iterations, where the first 2000 iterations are treated as burn-in, and with various numbers of factors, $Q=8, 10, 16$.
Second, the rolling-window method \citep{shumway2011} that partitions the time series into $B$ overlapping time blocks and estimates the local power spectrum by smoothed periodogram matrices with bandwidth chosen via generalized cross-validation (GCV)  \citep{ombao2011}. In our simulations, we consider three temporal block sizes, $B=64, 128, 256$.

We investigate the performance of an estimator of a spectral matrix $f(u,\omega)$ through the mean integrated squared error (MISE), which can be obtained by averaging squared errors across equally spaced time points and frequency grid as follows
\begin{equation*}
\text{MISE} = [T(K+1)]^{-1}\sum_{t=1}^T \sum_{k=0}^K \left |\left | \widehat{f}(t/T, \omega_{k}) - f(t/T, \omega_{k}) \right | \right |_F^2,
\end{equation*}
where $||\cdot||_F$ denotes the Frobenius norm of a matrix,  and $\widehat{f}(t/T, \omega_{k})$ is the estimated posterior mean of $f(t/T, \omega_{k})$. Moreover, we also consider the following quantities,
\begin{eqnarray*}
\text{MISE}_d &=& [T(K+1)]^{-1}\sum_{t=1}^T \sum_{k=0}^K \left |\left | \diag[\widehat{f}(t/T, \omega_{k})] - \diag[f(t/T, \omega_{k})] \right | \right |^2,  \\
\text{MISE}_o &=& [T(K+1)]^{-1}\sum_{t=1}^T \sum_{k=0}^K \left |\left | \widehat{f}(t/T, \omega_{k}) - f(t/T, \omega_{k}) \right | \right |_{F*}^2,
\end{eqnarray*}
where $||\cdot||_{F*}$ denotes the Frobenius norm, discarding the diagonal entries such that $||A||_{F*} = \sqrt{\sum_{i=1}^P \sum_{j \ne i} A_{ij}^2}$. In other words, $\text{MISE}_d$ and $\text{MISE}_o$ are the mean squared errors of the diagonal and off-diagonal elements, respectively.

\subsubsection{Piecewise Stationary Process}\label{subsec:abrupt}
\begin{figure}[p]
		\vspace*{-0.8in}
		  \begin{subfigure}
		  	\centering
		  	\includegraphics[width=16.4cm]{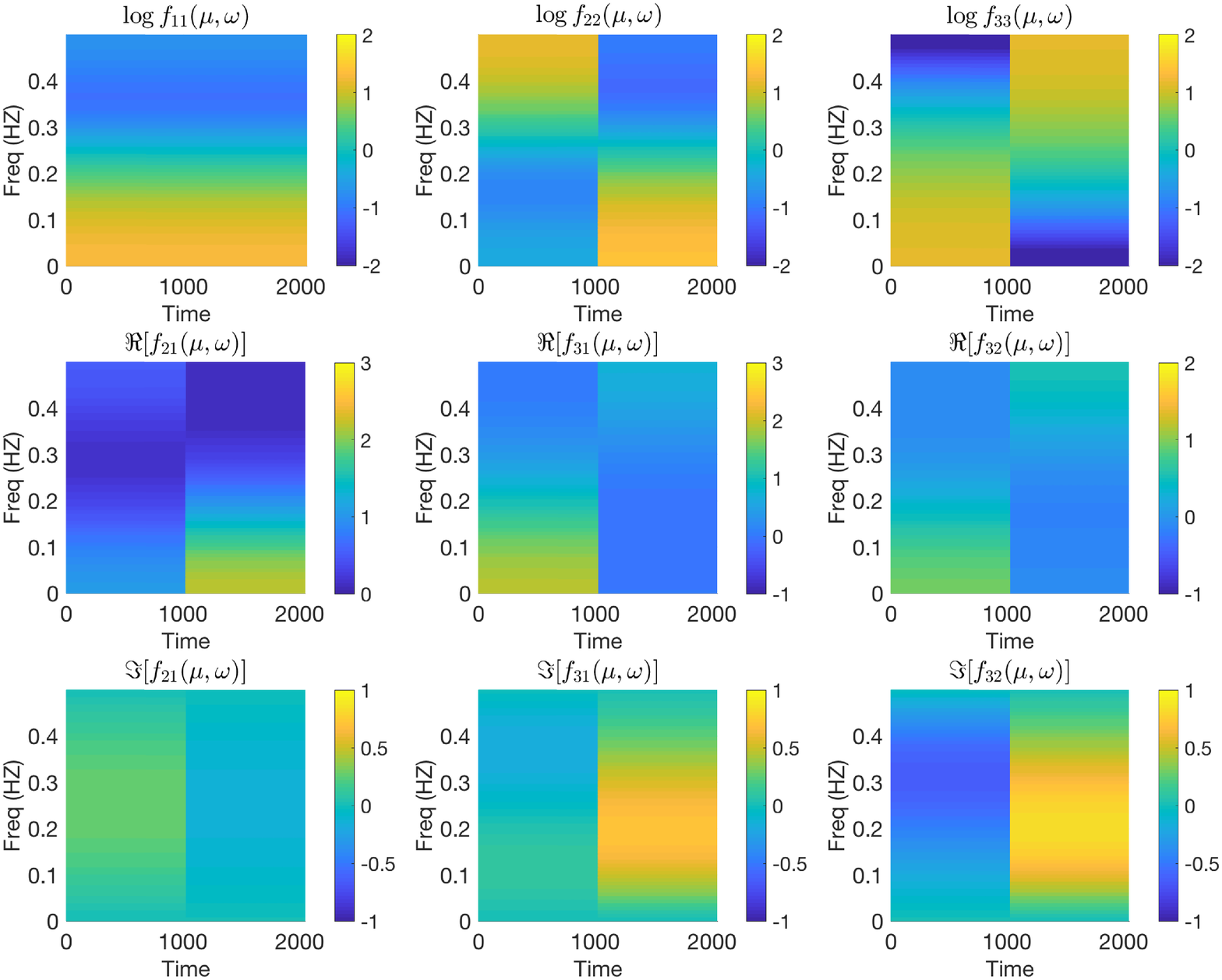}
		  \end{subfigure}
		  \begin{subfigure}
		  	\centering
		  	\includegraphics[width=16.4cm]{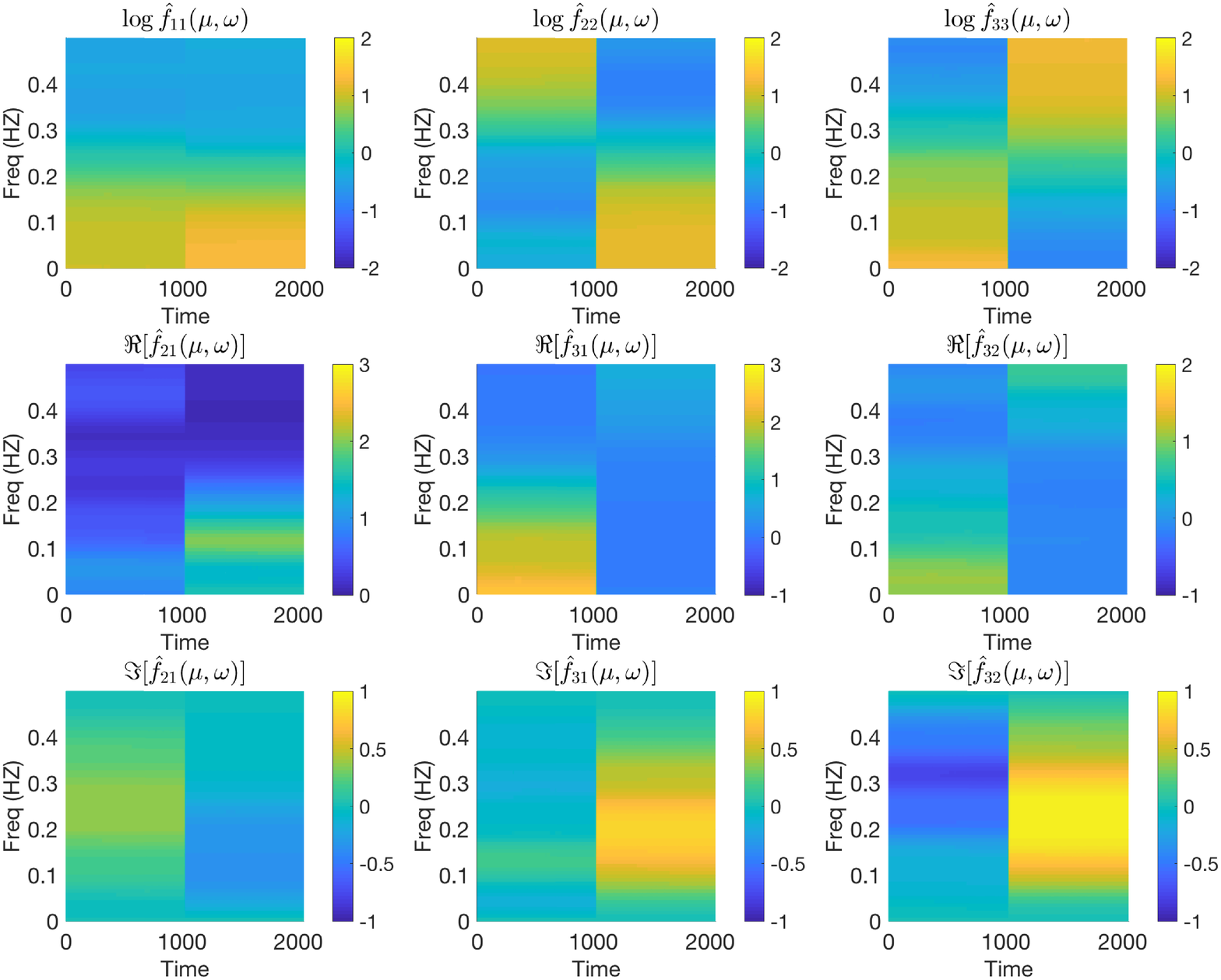}
		  \end{subfigure}
		\vspace{-1.1cm}
	\caption{Rows 1-3: the true time-varying log power spectra: $\log f_{11}(u, \omega)$, $\log f_{22}(u, \omega)$, and $\log f_{33}(u, \omega)$; and the real and imaginary parts of the cross-spectra: $f_{21}(u, \omega)$, $f_{31}(u, \omega)$, and $f_{32}(u, \omega)$ of process \eqref{eq:abrupt}. Rows 4-6: the corresponding estimates based on the proposed method.}
	\label{abrupt}
\end{figure}


We simulate 100 independent $P$-dimensional piecewise stationary time series of length $T=2048$ from
\begin{equation}\label{eq:abrupt}
X_t =
\begin{cases}
\mb \epsilon_{t} + \Phi_{11} \mb \epsilon_{t-1} + \Phi_{12} \mb\epsilon_{t-2}   & \quad \text{if }  1 \le t \le 1024\\
\mb \epsilon_{t} + \Phi_{21} \mb \epsilon_{t-1} + \Phi_{22} \mb \epsilon_{t-2}   & \quad \text{if } 1025 \le t \le 2048.
\end{cases}
\end{equation}
Each of the $P \times P$ coefficient matrices are block diagonal with $P/3$ blocks such that $\Phi_{11}= I_{P/3} \otimes \Phi_{11}^0$, $\Phi_{21}= I_{P/3} \otimes \Phi_{21}^0$, $\Phi_{12}= I_{P/3} \otimes \Phi_{12}^0$, and $\Phi_{22}= I_{P/3} \otimes \Phi_{22}^0$, where $I_{P/3}$ is a $P/3 \times P/3$ identity matrix, $\otimes$ denotes the kronecker product, and
\begin{equation*}
 \Phi_{11}^0 = \begin{pmatrix}
0.6 &~  0 &~ 0   \\
0.2 &~  -0.6 &~ 0  \\
0.1 & ~ 0.2 & 0.6
\end{pmatrix},
 \Phi_{21}^0 = \begin{pmatrix}
0.6 &~  0 &~ 0   \\
0.2 &~  0.6 &~ 0  \\
-0.1 & ~ -0.2 & -0.6
\end{pmatrix},
 \Phi_{12}^0 =\Phi_{22}^0 = \begin{pmatrix}
0.3 &~  0 &~ 0   \\
0 &~  -0.3 &~ 0  \\
0 & ~ 0 & 0
\end{pmatrix}.
\end{equation*}
The white noise $\mb \epsilon_t \stackrel{i.i.d}{\sim} N_p(\mb 0, \Omega)$ with $\Omega =I_{P/3} \otimes \Omega^0$, where $\Omega^0$ has 1's on the diagonal and 0.5 off the diagonal.
The true $P \times P$ time-varying power spectrum is  $f(u,\omega) = \Phi(u,\omega) \Omega \Phi(u, \omega)^*$ where  $\Phi(u,\omega) = I + \Phi_{11} \exp(-2 \pi i \omega) + \Phi_{12} \exp(-4 \pi i \omega)$ for $u \in [0, 1/2]$, $\Phi(u,\omega) = I - \Phi_{21} \exp(-2 \pi i \omega) + \Phi_{22} \exp(-4 \pi i \omega)$ for $u \in (1/2, 1]$ \citep[Chapter 9.4]{priestley1981}. Some of the true time-varying log power spectra and cross-spectra, and their estimates under the proposed procedure are displayed in Figures \ref{abrupt}, which shows that the true power spectrum changes abruptly at $t=1000$ and the proposed approach can accurately capture these dynamics.

\subsubsection{Slowly Varying Process}\label{subsec:slow}

\begin{figure}[p]
	\centering
		\includegraphics[width=16cm,height=11cm]{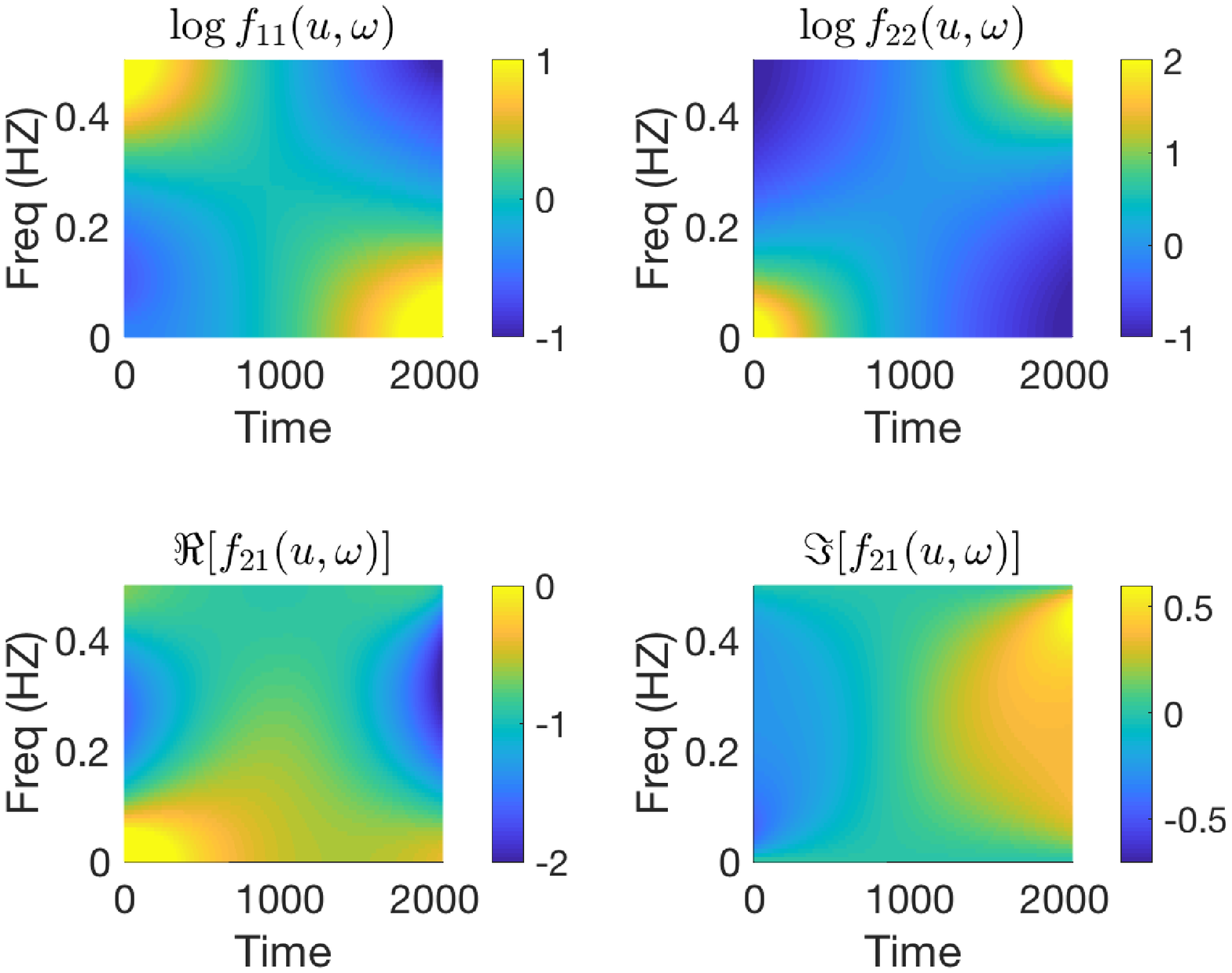}
		\includegraphics[width=16cm,height=11cm]{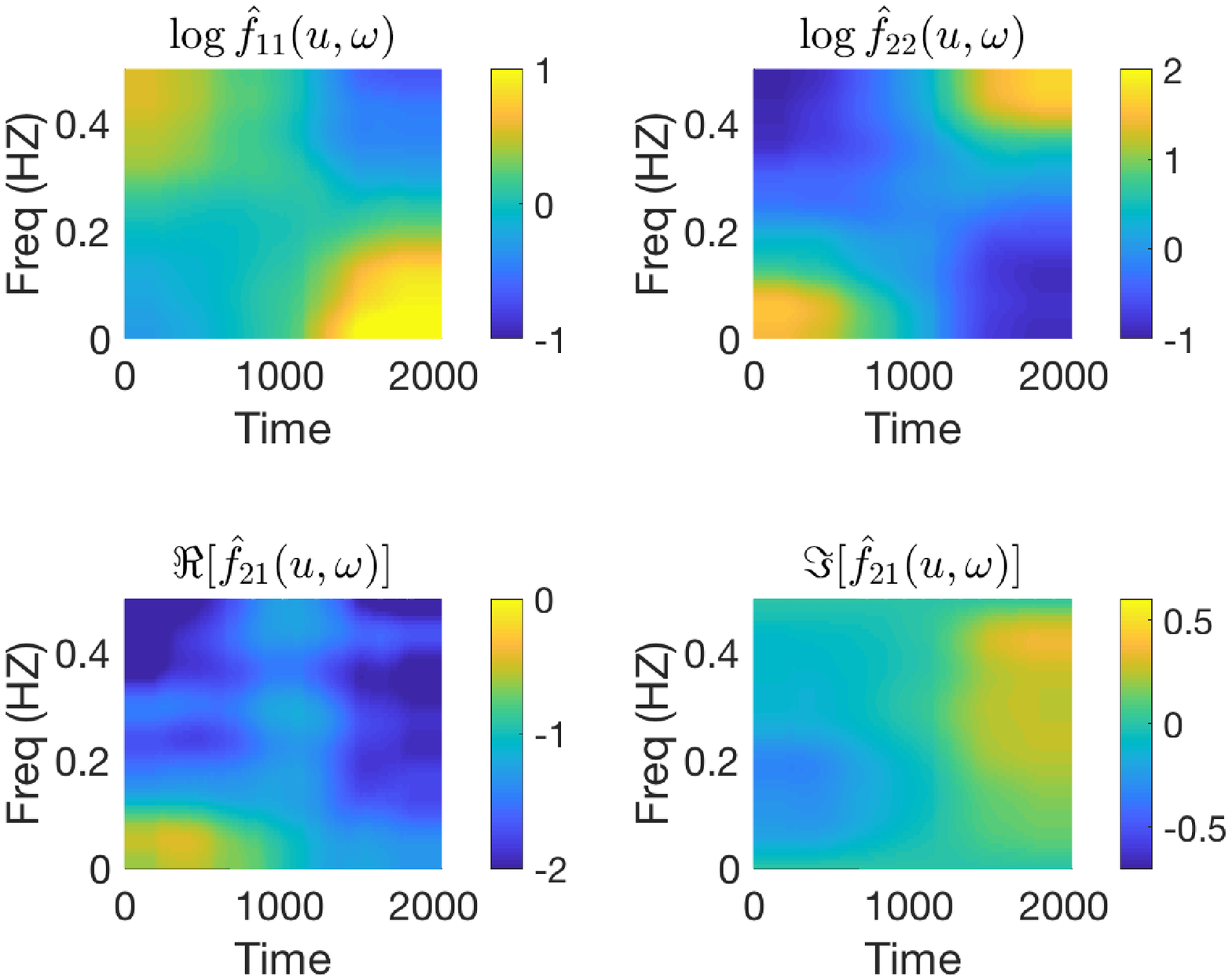}
	\caption{Rows 1-2: the true time-varying log power spectra: $\log f_{11}(u, \omega)$ and $\log f_{22}(u, \omega)$; and the real and imaginary parts of the cross-spectra $f_{21}(u, \omega)$ of process \eqref{eq:slow}. Rows 3-4: the corresponding estimates based on the proposed method.}
	\label{slow}
\end{figure}

We consider a slowly varying $P$-dimensional vector autoregressive process VAR(1) of length $T=2048$ and simulate 100 time series independently from
\begin{equation}\label{eq:slow}
\mb X_t = \Theta_{t} \mb X_{t-1} + \mb \epsilon_{t},
\end{equation}
where $\mb \epsilon_t \stackrel{i.i.d}{\sim} N_p(\mb 0, \Omega)$, $\Omega$ is as in Section \ref{subsec:abrupt}
and $\Theta_{t} = I_{P/2} \otimes \Theta^0_t$, with
  $$\Theta^0_t = \begin{pmatrix}
 	\theta^0_1(t) &~ 0.1   \\
 	0 &~ \theta^0_2(t)
 \end{pmatrix},
$$
and $\theta^0_1(t) = -0.5 + t/T$, $\theta^0_2(t)= 0.7 - 1.4t/T$ for $t=1, \cdots, T$. The true $P \times P$ time-varying power spectrum is  $f(t/T,\omega) = \Theta^{-1}(t/T,\omega) \Omega \Theta^{-1}(t/T, \omega)^*$ where  $\Theta(t/T,\omega) = I + \Theta_t \exp(-2 \pi i \omega)$  \citep[Chapter 9.4]{priestley1981}. Some of the true time-varying spectra and cross-spectra, and their estimates under the proposed procedure are displayed in Figure \ref{slow}, which shows that the true power spectrm changes smoothly over time and the proposed approach can successfully capture the slowly varying dynamics.

\subsubsection{Results}\label{subsec:results}

The mean and standard deviation of the three MISE criteria are presented in Table \ref{tab:MISE}. For both processes, the proposed method outperforms the rolling-window estimator, as the former has smaller MISE, $\text{MISE}_d$, and $\text{MISE}_o$ across both $P=24$ and $p=48$.  This is particularly true when estimating the off-diagonal elements as  $\text{MISE}_o$ values corresponding to the rolling-window method are 4-16 times larger than those corresponding to the proposed factor model.  Obtaining accurate estimates of the off-diagonal elements of a spectral matrix is crucial, since they can be used to derive many useful measures of between-series relationships, such as coherence and partial coherence, which provide valuable information for understanding many neurophysiological time series observed simultaneously. Moreover, it should be noted that the proposed method has similar MISE values across the selected number of factors $Q$. This can be attributed to using the  multiplicative gamma process shrinkage priors, which reduce the sensitivity to the number of factors. 

\begin{table}[h]
	\caption {\label{tab:MISE} Simulation results for the piecewise stationary and slowly varying processes. Based on 100 repetitions, means (standard deviation) of MISE, $\text{MISE}_d$, and $\text{MISE}_o$ obtained through the rolling-window and the proposed factor model. }
	\vspace{0.3cm}
	\centering
	\scalebox{0.8}{
		\begin{tabular}{lllrrrrrrr}
			\hline
			Settings &  $P$  & ~  &  \multicolumn{3}{c}{Rolling-window} & ~& \multicolumn{3}{c}{Factor Model}    \\
			\cline{4-6} \cline{8-10}
			& ~ & & $B=64$ & $B=128$ & $B=256$ & ~ & $Q=8$ &$Q=10$ & $Q=16$                       \\    \hline
			Process \eqref{eq:abrupt}:
			& 24 & $\text{MISE}$  & 23.75 (1.01)   &  14.73 (0.80)   & 9.29 (0.52)   & ~  &   4.24 (0.31)  & 3.33 (0.11)  &3.27 (0.09) \\
			 Piecewise&     & $\text{MISE}_d$     & 2.54 (0.17) &  1.95  (0.14)     & 1.91 (0.09)  & ~   &  1.08 (0.06)   & 0.31 (0.04) &0.24 (0.06) \\
			\vspace{0.2cm}
			&  & $\text{MISE}_o$     & 21.21  (0.90)   & 12.78  (0.73)     &  7.38 (0.47) & ~   &  3.16 (0.26)   & 3.02 (0.11) &3.02 (0.06)  \\
			& 48 & $\text{MISE}$  & 92.42 (2.65)  &  56.19 (2.26)    & 33.33 (1.58)    & ~   & 10.00 (0.36)    &  9.31 (0.12) &   9.29 (0.14)\\
			&  & $\text{MISE}_d$     & 5.07 (0.23)     &  3.89 (0.17)   &  3.81 (0.10) & ~   &  3.61 (0.07)   &  2.79 (0.06) &  2.81 (0.05) \\
			&  & $\text{MISE}_o$     & 87.35 (2.53)   &   52.30 (2.17)    &  29.52 (1.53) & ~   &  6.39 (0.34)   &  6.35 (0.22) & 6.32 (0.21) \\
			\hline
			Process \eqref{eq:slow}:
			& 24 & $\text{MISE}$   & 23.17 (2.58)   &  17.23 (1.74)    & 14.91 (1.17)    & ~   &  4.04 (0.06)   &  3.98 (0.06)  &  3.97 (0.06) \\
			Slowly-varying &      & $\text{MISE}_d$  & 6.13   (0.37)   &  6.02  (0.30)   & 6.23 (0.31)   & ~   &  1.72 (0.02)   &    1.69 (0.02)   & 1.70 (0.03)\\
			\vspace{0.2cm}
			&      & $\text{MISE}_o$  & 17.04  (2.26)   & 11.21  (1.50)   & 7.68 (0.99) & ~   &   2.32 (0.05)  &  2.29 (0.04)  & 2.27 (0.02) \\
			& 48 & $\text{MISE}$   & 76.92 (7.00)    & 52.48 (4.99)  & 37.79 (3.27)   & ~   &  8.46 (0.12)   &  8.15 (0.08)  &  8.15 (0.10)  \\
			&  & $\text{MISE}_d$  	   &  12.30 (0.64)   & 12.03 (0.47)   & 13.04 (0.38)   & ~   &  4.24 (0.05)   &  4.12 (0.03) &  4.13 (0.04)  \\
			&  & $\text{MISE}_o$      &  64.63 (6.45)  & 40.56 (4.61)   & 24.75 (3.00) & ~   &   4.22 (0.09)    &  4.03 (0.06) & 4.02 (0.07) \\
			\hline
		\end{tabular}
	}
\end{table}

\subsection{Estimation of Partitions}\label{subsec:partitions}

In this section, we focus on evaluating the frequentist properties of the proposed method in estimating the number and location of partition points. First, to demonstrate that the proposed method can correctly identify a stationary process, we use the first piece of process \eqref{eq:abrupt}, that is, we simulate one hundred 48-dimensional time series of length $T=1024$ independently from
\begin{equation}\label{m3}
\mb X_t = \mb \epsilon_{t} + \Phi_{11} \mb \epsilon_{t-1} + \Phi_{12} \mb\epsilon_{t-2},
\end{equation}
where  $\Phi_{11}$,  $\Phi_{12}$, and $\mb \epsilon_t$ are as in Section \ref{subsec:abrupt}. Second, we simulate one hundred 48-dimensional time series independently from
\begin{equation}\label{m4}
 \mb X_t =
\begin{cases}
\mb \epsilon_{t}^{(1)} + \Phi_{11} \mb \epsilon_{t-1}^{(1)} + \Phi_{12}  \mb \epsilon_{t-2}^{(1)} &  \quad \text{if }  1 \le t \le 500,\\
\mb \epsilon_{t}^{(1)} + \Phi_{21} \mb \epsilon_{t-1}^{(1)} + \Phi_{22}  \mb \epsilon_{t-2}^{(1)} & \quad \text{if } 501 \le t \le 1000, \\
\mb \epsilon_{t}^{(2)} + \Phi_{31} \mb \epsilon_{t-1}^{(2)} + \Phi_{32}  \mb \epsilon_{t-2}^{(2)}   & \quad \text{if } 1001 \le t \le 2000,\\
\mb \epsilon_{t}^{(2)} + \Phi_{41} \mb \epsilon_{t-1}^{(2)} + \Phi_{42}  \mb \epsilon_{t-2}^{(2)}  & \quad \text{if } 2001 \le t \le 4000,
\end{cases}
\end{equation}
where $\Phi_{11}$,  $\Phi_{12}$, $\Phi_{21}$,  and $\Phi_{22}$ are  as in Section \ref{subsec:abrupt}, and $\Phi_{31}=\Phi_{21}$, $\Phi_{32} = \Phi_{22}$. We define $\Phi_{41}$ and $\Phi_{42}$ as block diagonal matrices such that $\Phi_{41}= I_{P/3} \otimes \Phi_{41}^0$ and $\Phi_{42}= I_{P/3} \otimes \Phi_{42}^0$, where
\begin{equation*}
\Phi_{11}^0 = \begin{pmatrix}
1.32 &~  0 &~ 0   \\
0.2 &~  -0.6 &~ 0  \\
0.1 & ~ 0.2 & 0.6
\end{pmatrix},
\Phi_{42}^0 = \begin{pmatrix}
-0.81 &~  0 &~ 0   \\
0 &~  -0.3 &~ 0  \\
0 & ~ 0 & 0
\end{pmatrix}.
\end{equation*}
The white noise terms $ \mb \epsilon^{(1)}_{t}$ , $\mb \epsilon^{(2)}_{t}$ are independent zero-mean $P$-dimensional Gaussian random variables whose covariance matrices have unit variances and pairwise correlations of $0.5$ and $0.9$, respectively. This process has both short segments (with 500 observations) and long ones (with 2000 observations),  and the change, at $t = 1000$,  in the off-diagonal elements of the covariance of the error term is subtle compared to the other changes.

\begin{table}[h]
	\caption {\label{tab:multi} Simulation results for processes \eqref{m3} and \eqref{m4} based on 100 repetitions: means (standard deviations) of the posterior probabilities $\Pr(L|\mb X)$.}
	\vspace{0.3cm}
	\centering
	\scalebox{1}{
		\begin{tabular}{lccccccccc}
			\cline{1-9}
			Settings & \multicolumn{8}{c}{Number of segments $L$} &    \\
			\cline{2-9}
			&  $L=1$ & $L=2$ & $L=3$ & $L=4$ & $L=5$ & $L=6$ & $L=7$& $L=8$                     \\    			\cline{1-9}
			Process \eqref{m3}  & 0.940    &  0.057  & 0.030   & 0  & 0   & 0   & 0   & 0 \\
			&(0.079)   & (0.073)   &  (0.007)   & (0)  & (0)  & (0)  & (0)  & (0)\\
			Process \eqref{m4} & 0 &  0    & 0.051  & 0.889 & 0.060 & 0 & 0 & 0 \\
			& (0) &  (0)    & (0.182)  & (0.194) & (0.103) & (0) & (0) & (0) \\
			\cline{1-9}
		\end{tabular}
	}
\end{table}

The proposed method was run using $Q=10$ factors for 10,000 iterations with a burn-in of 2,000.  Table  \ref{tab:multi} reports the mean and standard deviation of the estimated posterior probabilities of the numbers of segments, $\Pr(L|\mb X)$. For process \eqref{m3}, the proposed method correctly identifies a stationary process with a high probability:  $\widehat{\pr}(L=1|\mb X) = 0.940$. For process \eqref{m4}, the proposed method also correctly assigns the highest posterior probability to $L=4$, i.e., $\widehat{\Pr}(L = 4|X) = 0.889$.


\section{Analysis of TMS-Evoked High-Density EEG}\label{sec:eeg}

Abnormal neurophysiological activity is consistently observed in patients with schizophrenia and is associated with poor cognitive ability. To examine biological correlates, identify potential biomarkers, and to guide individualized treatment of mental illness, physicians often use hdEEG to measure electrophysiological activity simultaneously across multiple regions, or channels, of the brain. Transcranial magnetic stimulation (TMS) is a noninvasive method of brain stimulation that uses an insulated coil placed over a chosen location of the scalp to induce a magnetic field that excites cells in the cortex of the brain.  Concurrent TMS and hdEEG are used to record TMS-evoked local and general brain activity with millisecond precision.  Time-varying spectral analysis of hdEEG during TMS provides important neurobiological information that can be used to assess and guide treatments of psychiatric disorders \citep{kaskie2018}.

We demonstrate the proposed method through the analysis of hdEEG during TMS from a patient during hospitalization for their first psychotic episode.
The epoch considered was taken from 1000 milliseconds before and 1000 milliseconds after a TMS delivery to the primary motor cortical area of the patient's brain with a sampling rate of 5000 samples per second (5000 Hz).  Data were preprocessed using the \texttt{TMSEEG} Matlab GUI \citep{atluri2016} where signals were downsampled to 1000 Hz. Independent component analysis (ICA) was used to remove pulse and ripple artifacts, IIR bandpass (1-80 Hz) and notch filters (60 Hz) were applied, and signals were referenced and standardized to unit variance.  The resulting data, which are displayed in Figure \ref{tseeg}, are a 64-dimensional time series, i.e., $P=64$, of length $T = 2000$.

The proposed method was fit with $Q=15$ factors, and run for 10,000 iterations with burn-in of 5,000. The goal of our study is to analyze the entire time-varying spectral matrix, $f(u, \omega)$, as well as certain functions of the power spectrum. Specifically, three types of quantities are of interest. First, the time-varying power spectra $f_{j j}(u, \omega), ~j=1, \cdots, P$ and the pairwise squared coherences
$$\rho^2_{jk}(u,\omega) = |f_{jk}(u,\omega)|^2/[f_{jj}(u,\omega) f_{kk}(u,\omega)], ~j,k=1,\cdots, P,$$
can be easily estimated. Second, the proposed adaptive Bayesian method allows one to conduct inference on frequency-collapsed functionals. In particular, power within several frequency bands, including the gamma band (31-80 HZ), contains important neurophysiological information \citep{ferrarelli2018}. Frequency-band collapsed measures can be computed as integrals of the power spectra. For example, the time-varying gamma-band collapsed spectral matrices are given by
\begin{eqnarray}\label{gamma}
	f^{\gamma}(u) &=& \int_{31}^{80} f(u, \omega) d\omega.
\end{eqnarray}
Third, the time-varying frequency-collapsed squared coherences can be derived to investigate connectivity between brain regions across time at frequency bands of interest. A commonly used measure of local connectivity is the beta-band (16-31 HZ) coherences. The beta-band squared coherence between channels $j$ and $k$ is defined as
\begin{eqnarray*}
	\rho_{jk}^{2,\beta} (u)&=& \left | \int_{16}^{31} f_{jk} (u, \omega)  d \omega \right |^2  / \left \{  f_{jj}^{\beta}(u) f_{kk}^{\beta}(u)  \right \},
\end{eqnarray*}
where $f^{\beta}_{j j}(u) =\int_{16}^{31} f(u, \omega) d\omega$ is the time-varying beta-band power of channel $j$. The following sections provide the results of our analyses of the TMS-evoked hdEEG.

\subsection{Time-Varying Power Spectra and Squared Coherences}

\begin{figure}[p]
	\centering
	\begin{minipage}[b]{1\textwidth}
		\includegraphics[width=\textwidth]{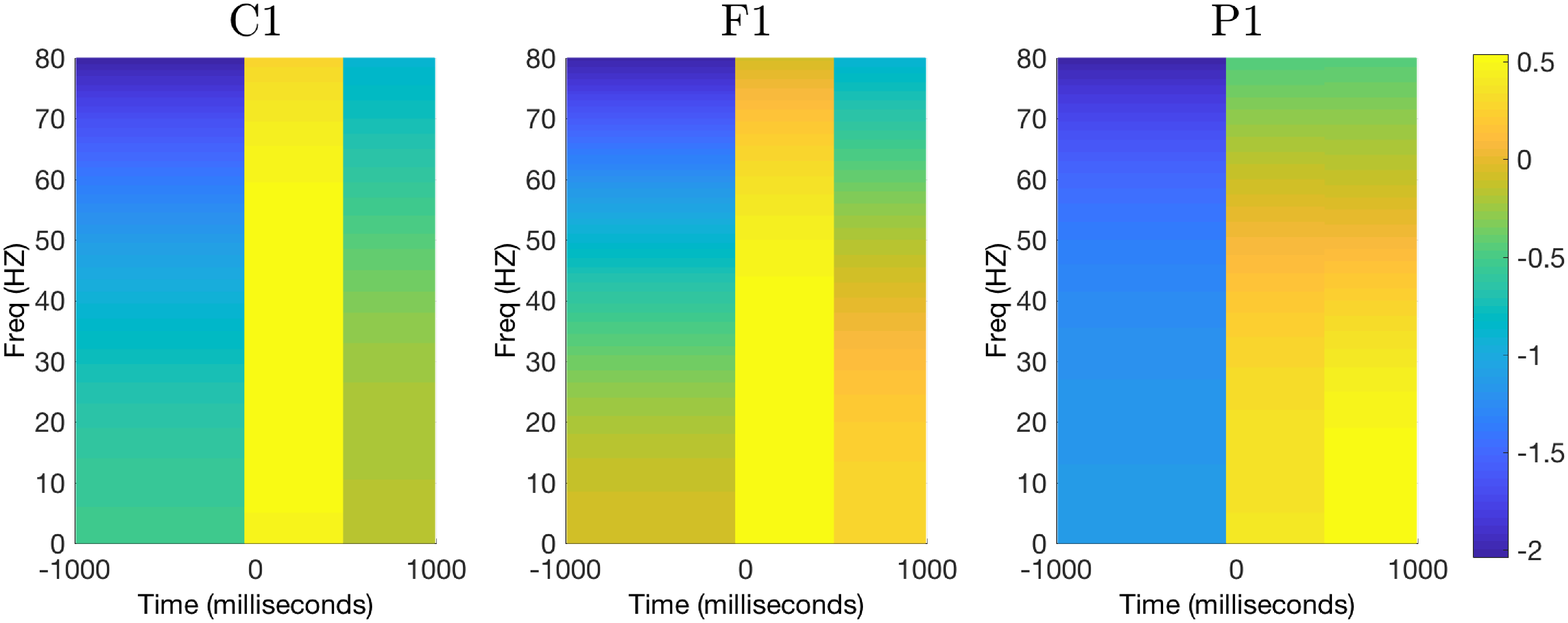}
	\end{minipage}
	\begin{minipage}[b]{1\textwidth}
		\includegraphics[width=\textwidth]{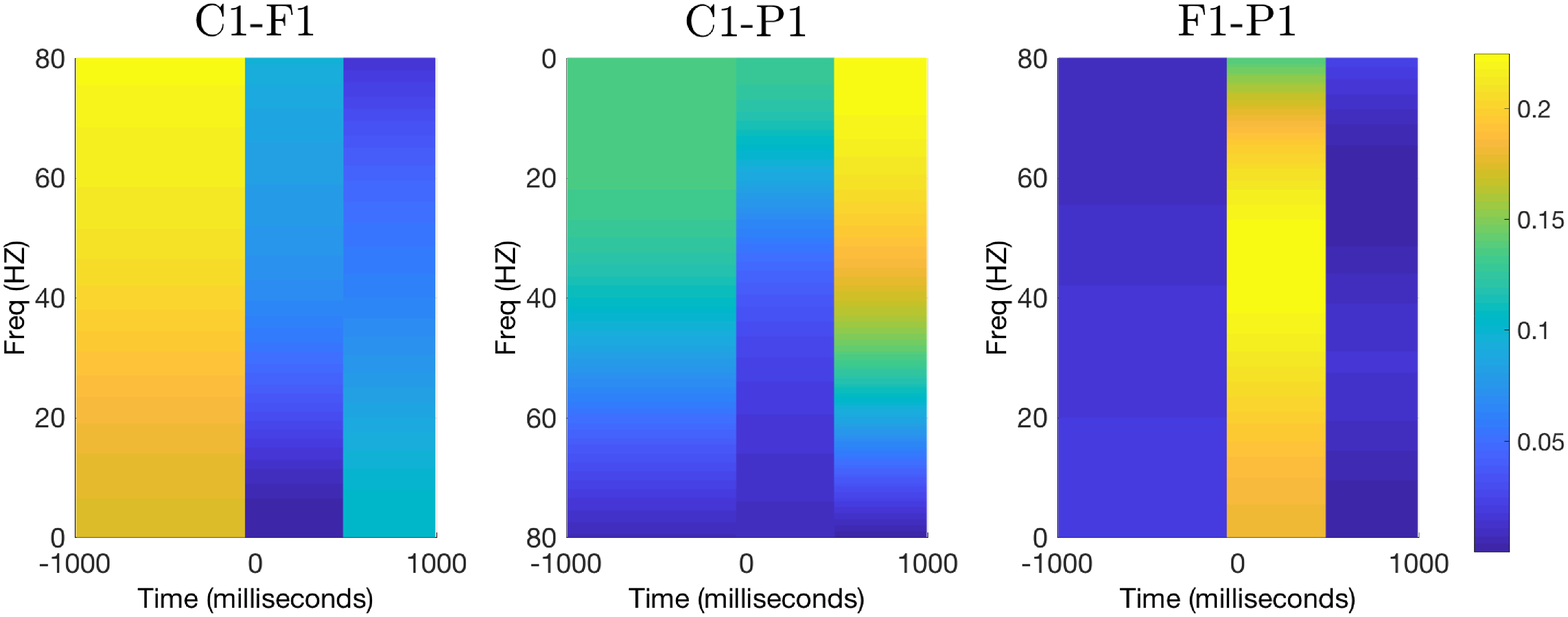}
	\end{minipage}	
	\caption{Estimates of the time-varying log spectra (first row) and pairwise squared coherences (second row) of the TMS evoked-EEG in channels C1, F1, and P1.}
	\label{spectTMS}
\end{figure}

\begin{figure}[p]
	\centering
	\begin{minipage}[b]{1\textwidth}
		\includegraphics[width=\textwidth]{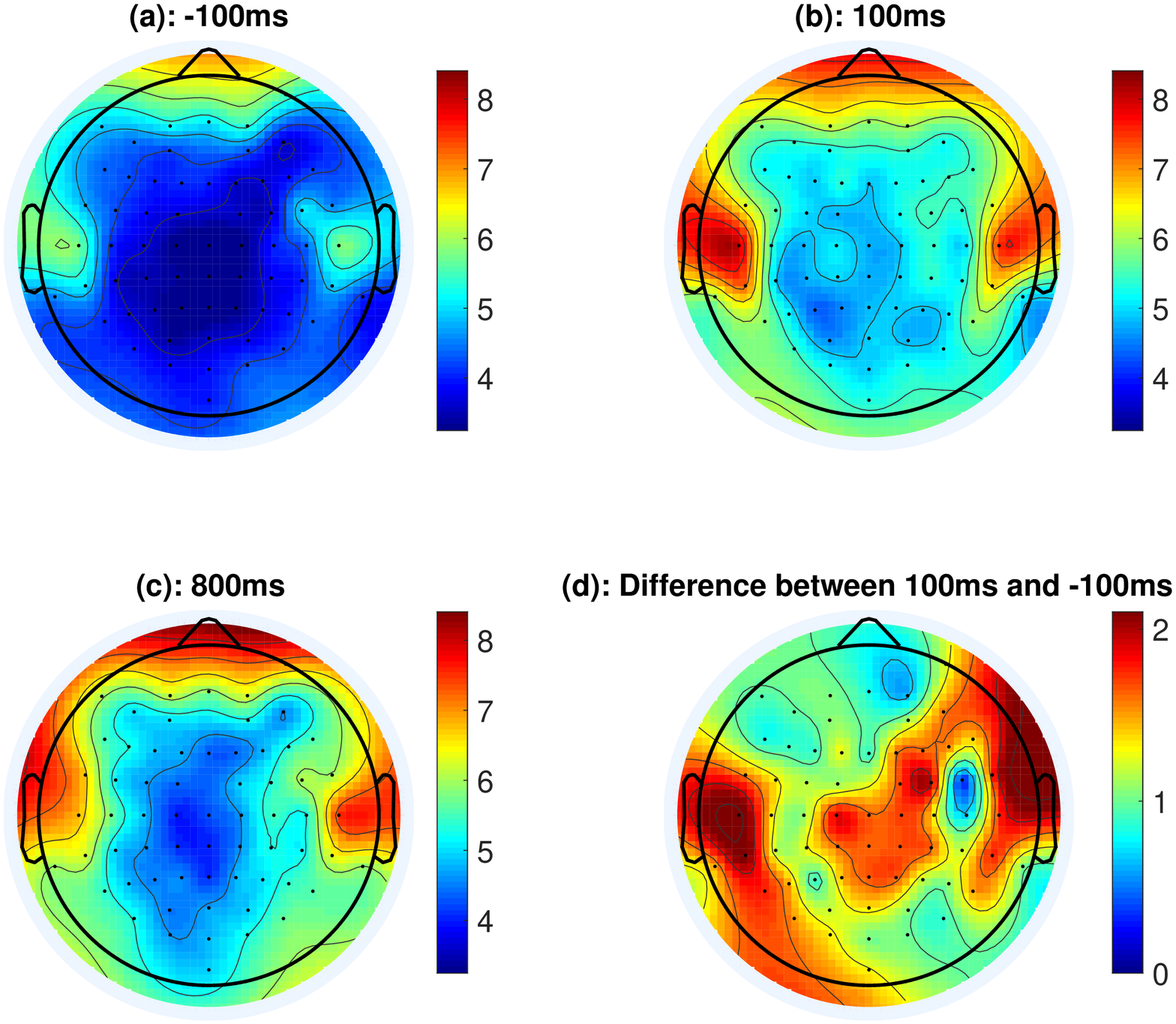}
	\end{minipage}
	\caption{The estimated gamma-band log power of 64-channel hdEEG at time points: (a) -100 ms, (b) 100 ms, and (c) 800 ms; and (d) the difference of gamma band log power between time point 100 ms and -100 ms.  }
	\label{spectGamma}
\end{figure}

Figure \ref{spectTMS} presents the estimated time-varying log-power spectra and  the corresponding pairwise squared coherences from three channels: C1, F1, and P1, which are located in the primary motor cortex, the frontal cortex, and the parietal cortex, respectively. It is clear that the log-power spectra and squared coherences changed abruptly, resulting in three approximately stationary segments, which can be identified as three periods: before TMS (the baseline EEG), immediately after TMS, and recovering from TMS. In general, our results indicate that the estimated power spectra increased at all frequencies after TMS delivery. It should be noted that channel C1, which is located in the primary motor cortex (the region where TMS was induced), had a greater power increase compared to channels F1 and P1.  The squared coherence between C1 and F1 and between C1 and P1 decreased dramatically at all frequencies after TMS delivery, indicating that the stimulation restricts the correlation between the primary motor cortical area and other regions of the brain. Interestingly, channels F1 and P1, which are located far apart, experienced an increase in the squared coherence after the TMS delivery. This calls for further research on how channels that are not in actively stimulated regions behave after TMS, which has not been explored in the biomedical literature.

\subsection{Time-Varying Frequency-Collapsed Power Spectra}

Figure \ref{spectGamma} (a)-(c) presents topoplots of the estimated gamma-band collapsed spectral matrices defined in \eqref{gamma} at different time points: $t=-100, 100, ~ \text{and} ~800$ milliseconds (ms).  Comparing Figure \ref{spectGamma} (a) and (b), we see that the log gamma-band power increased drastically after the TMS delivery ($t=100$ ms) for all channels. However, as indicated by Figure \ref{spectGamma} (d), which displays the differences of log gamma-band power between time points 100 ms and -100 ms, the amount of increases of the gamma-band power in the primary motor cortex is larger than that in the other areas. Moreover, the prefrontal cortex had the smallest amount of increase in gamma-band power. It should be noted that similar characteristics have been observed in patients with schizophrenia \citep{ferrarelli2008}.  Our findings suggest that our patient, who has experienced a first psychotic episode but does not yet meet the clinical requirement for a diagnosis of schizophrenia,  exhibits some neurophysiological characteristics that are similar to those in patients with schizophrenia compared to healthy controls, which could potentially serve as a subclinical biomarker.

\subsection{Time-varying Frequency-Collapsed Coherences}

\begin{figure}[h]
	\centering
	\includegraphics[width=1\textwidth]{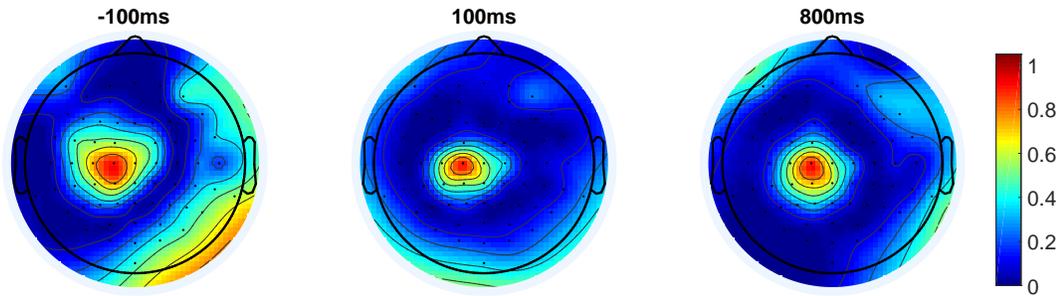}
	\caption{The estimates beta-band squared coherences of C1 with respect to all other channels at different time points}
	\label{beta_coh}
\end{figure}

\begin{figure}[h]
	\centering
	\includegraphics[width=1\textwidth]{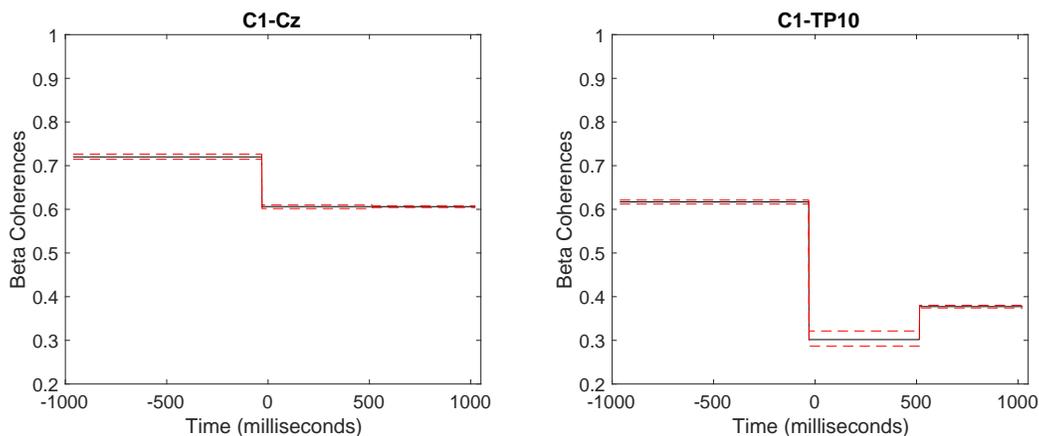}
	\caption{The estimated beta-band squared coherences of C1 with respect to Cz (left) and TP10 (right) over time  with 95\% pointwise credible intervals. }
	\label{tp10}
\end{figure}

Figure \ref{beta_coh} displays the estimated beta-band squared coherence of C1 with respect to all other channels at time points: $t=-100, 100, ~ \text{and} ~800$ ms. Broadly, during the baseline period ($t=-100$ ms), C1 is connected to the channels that are also within the primary motor cortex, as well as to channels within the right hemisphere of the parietal and frontal cortexes.  However, after the TMS ($t=100$ ms) delivery, C1 only retains high beta-band coherences within the primary motor cortex. Beta-band connectivity between C1 and other brain regions decreases or disappears.  When $t=800$ ms, the connectivity between C1 and the right hemisphere of the parietal and frontal cortexes begin to be restored. This phenomenon can be further illustrated in Figure \ref{tp10}, which shows the estimated time-varying beta squared coherences between C1 and TP10, and C1 and Cz along with 95\% credible intervals (as the 2.5 and 97.5 empirical percentiles of the MCMC iterates after the burn-in period). It should be noted that
Cz is located close to C1, while TP10 is located at the right hemisphere of the parietal cortical area.

\section{Final Remarks} \label{remarks}
This article introduces the first adaptive approach to analyzing the time-varying power spectrum of a high-dimensional nonstationary time series. A novel frequency-domain factor model is developed using a prior that is the tensor product of penalized splines and  multiplicative gamma process priors. A scalable MCMC algorithm is developed for model fitting, allowing for inference on both abrupt and slowly varying processes. We conclude this article by discussing some limitations and related future extensions. First, the proposed procedure is designed for the analysis of a single high-dimensional time series. However, in many applications, interest lies in  the analysis of replicated high-dimensional time series in order to understand how time-varying spectra are associated with other covariates, such as clinical symptoms. A possible extension of the proposed method for the analysis of replicated time series could involve adaptively dividing the grid of time and covariate values into approximately stationary blocks in a manner similar to that proposed in the univariate setting by \cite{bruce2018}.  Second, our analysis of high-dimensional time series assumes that all time series  have the same sampling rate. However, time series can be sampled at different rates. For example, one could be interested in the joint spectral analysis of other physiological signals, such as heart rate variability, which is typically sampled around 1 Hz, with hdEEG.  Future research will focus on developing adaptive spectral analysis of high-dimensional time series with different sampling rates. Finally, the use of a tensor product prior allows us to obtain a good approximation to the locally stationary time series and to reduce the sensitivity to the number of factors. However, one could automatically select the number of factors by adopting an adaptive truncation approach akin to that used by \cite{Bhattacharya2011}.

\bibliographystyle{asa}
\bibliography{factorlib.blb}

\newpage
	\begin{center}
		{\LARGE\bf Supplementary Materials to ``Adaptive Bayesian Spectral Analysis of High-dimensional Nonstationary Time Series"}
	\end{center}

\setlength{\baselineskip}{22pt}  

\begin{center}
	\section*{Abstract}
\end{center}

This document contains supplemental materials to the article ``Adaptive Bayesian Spectral Analysis of High-dimensional Nonstationary Time Series." Section 1 presents  the details of the sampling scheme for spectral anlysis of a {\it stationary} high-dimensional time series. Section 2 discusses the technical details of the SAMC sampling scheme that was outlined in Section 3.2 of the main manuscript.



\setcounter{section}{0}
\section{Sampling Scheme for Stationary Time Series}

After initializing all the parameters, each iteration of the sampling scheme consists of the following steps.
\begin{enumerate}
	\item Sample factors $\mb D_{k}$, for $k=1, \cdots K$, independently from the conditional posterior distributions
	\begin{equation}\label{D}
	p(\mb D_{k} \mid \cdots) \sim CN (\mb \mu_{D_{k}}, \Sigma_{D_{k}}),
	\end{equation}
	where $\Sigma_{D_{k}} = (\Lambda_{k}^* \Sigma_{\epsilon}^{-1} \Lambda_{k} + I_{Q})^{-1}$ and $\mb \mu_{D_{k }} = \Sigma_{D_{k }} \Lambda_{k}^* \Sigma_{\epsilon}^{-1} \mb Y_{k }$.
	
	\item Let $\Lambda_{k}^{(pq)}$ be the $pq$th entry of $\Lambda_{k}$, $\mb \alpha_{pq} = (\alpha_{p q 0}, \cdots \alpha_{p q S-1})'$ and $\mb \beta_{pq } = (\beta_{p q  1}, \cdots \beta_{p q S})'$ be vectors of coefficients of basis functions defined in \eqref{one} and \eqref{two} of the main document, and $W^{(re)}$ and $W^{(im)}$ be design matrices with corresponding coefficients $\mb \alpha_{pq}$ and $\mb \beta_{pq }$ such that $\mb \Lambda^{(pq)} = (\Lambda_{1}^{(pq)}, \cdots, \Lambda_{K}^{(pq)} )' = W^{(re)} \mb \alpha_{pq} + W^{(im)} i \mb \beta_{pq}$. We sample  $\mb \alpha_{pq}$ and $\mb \beta_{pq}$, $p=1, \cdots P$, $q=1, \cdots, Q$, independently from the conditional posterior distributions, respectively,
	\begin{eqnarray}
	\label{RE}
	p(\mb \alpha_{pq } \mid \cdots) &\sim& N (\mb \mu_{\alpha_{pq}}, \Sigma_{\alpha_{pq}}), \\
	\label{IM}
	p(\mb \beta_{pq} \mid \cdots) &\sim& N (\mb \mu_{\beta_{pq}}, \Sigma_{\beta_{pq}})
	\end{eqnarray}
	and then update $\mb \Lambda^{(pq)}  = W^{(re)} \mb \alpha_{pq} + i W^{(im)}  \mb \beta_{pq}$. The exact form of the conditional means, $\mb \mu_{\alpha_{pq}}$ and $\mb \mu_{\beta_{pq}}$, and covariances,  $\Sigma_{\alpha_{pq}}$ and $\Sigma_{\beta_{pq}}$ are 	$$\Sigma_{\alpha_{pq }} = \left [  \Omega_{p q }^{(re)^{-1}} + 2\sigma_{\epsilon,p}^{-2} \sum_{k=1}^K |D_{k }^{(q)}|^2 \mb \omega_k^{(re)} \mb \omega_k^{(re)'} \right ]^{-1},$$
	$$\mb \mu_{\alpha_{pq}} = \frac{2}{\sigma_{\epsilon,p}^2} \Sigma_{\alpha_{pq}} \left [  \sum_{k=1}^K \Re\{ - Y_{k  p}^* D_{k}^{(q)} + S_{k }^{(p+\backslash q)} D_{k }^{(q)*}  \} \mb \omega_k^{(re)}    \right ], $$
	$$\Sigma_{\beta_{pq }} = \left [  \Omega_{p q}^{(im)^{-1}}+ 2\sigma_{\epsilon,p}^{-2} \sum_{k=1}^K |D_{k}^{(q)}|^2 \mb \omega_k^{(im)} \mb \omega_k^{(im)'} \right ]^{-1},$$
	$$\mb \mu_{\beta_{pq }} = -\frac{2}{\sigma_{\epsilon,p}^2} \Sigma_{\beta_{pq}} \left [  \sum_{k=1}^K \Im\{ - Y_{k p}^* D_{k}^{(q)} + S_{k}^{(p+\backslash q)}  D_{k}^{(q)*}  \}  \mb \omega_k^{(im)}    \right ], $$
	where	
	\begin{eqnarray*}
		\Omega_{p q}^{(re)} &=& \diag\left \{ \psi _{q, (re)}^{-1}, (2 \pi)^{-1} \tau_{p q, (re)}^2 \psi _{q, (re)}^{-1}, \cdots, [2 \pi (S-1)]^{-1} \tau_{p q, (re)}^2 \psi _{q, (re)}^{-1} \right \}, \\
		\Omega_{p q }^{(im)} &=& \diag\left \{(2 \pi)^{-1} \tau_{p q , (im)}^2 \psi _{q , (im)}^{-1}, \cdots, (2 \pi S)^{-1} \tau_{p q, (im)}^2 \psi _{q, (im)}^{-1} \right \},
	\end{eqnarray*}	
	and $\mb \omega_k^{(re)}$ and $\mb \omega_k^{(im)}$ are the $k$th row of $W^{(re)}$ and $W^{(im)}$, respectively, $S_{k}^{(p+\backslash q)} = \sum_{h \in \mathcal{Z}_Q \backslash q} \Lambda_{k}^{(ph)} D_{k}^{(h)}$, and $\mathcal{Z}_Q = \{1,2,\cdots,Q  \}$.

	\item Sample the smoothing parameters $\tau_{p q, (re)}^2$, $\tau_{p q, (im)}^2$, $p=1, \cdots P$, $q=1, \cdots, Q$, from the conditional posterior distributions
	\begin{eqnarray}
	\label{tau1}
	p(\tau_{p q, (re)}^2 \mid \cdots) &\sim& IG \left \{ \frac{(S+\nu - 1)}{2}, \frac{\mb \alpha_{pq}'\mb \alpha_{pq}}{2 \psi_{q,(re)}^{-1}} + \frac{\nu}{g_{pq, (re)}} \right \}, \\
	\label{tau2}
	p(\tau_{p q, (im)}^2 \mid \cdots) &\sim& IG \left \{ \frac{(S+\nu)}{2}, \frac{\mb \beta_{pq}'\mb \beta_{pq }}{2 \psi_{q,(im)}^{-1}} + \frac{\nu}{g_{pq, (im)}} \right \},
	\end{eqnarray}
	respectively. Then, we update the hyperparameters $g_{pq,(re)}$, $g_{pq,(im)}$ by sampling from
	\begin{eqnarray}
	\label{g1}
	p(g_{p q , (re)} \mid \cdots) &\sim& IG \{(\nu+ 1)/2, \nu/\tau_{p q , (re)}^2 + 1/G_{\tau}^2\}, \\
	\label{g2}
	p(g_{p q , (im)} \mid \cdots) &\sim& IG \{(\nu+1)/2,  \nu/\tau_{p q, (im)}^2 + 1/G_{\tau}^2\}.
	\end{eqnarray}
	
	\item Sample the shrinkage parameters $\phi_{1, (re)}$ and $\phi_{1, (im)}$, from the conditional posterior distributions
	\begin{eqnarray*}
		p(\phi_{1, (re)} \mid \cdots) &\sim& Ga \left \{ a_1 +\frac{PQS}{2}, \sum_{q=1}^{Q} \psi_{q,(re)}^{(1)} \sum_{p=1}^{P} \frac{\mb \alpha_{pq}'\mb \alpha_{pq}}{2 \tau_{pq, (re)}^2}\right \} ,\\
		p(\phi_{h, (im)} \mid \cdots) &\sim& Ga \left \{ a_1 +\frac{PQS}{2},  \sum_{q=1}^{Q} \psi_{q, (im)}^{(1)} \sum_{p=1}^{P} \frac{\mb \beta_{pq}'\mb \beta_{pq}}{2 \tau_{pq, (im)}^2}\right \},
	\end{eqnarray*}
	and for $h\ge 2$, sample $\phi_{h, (re)}$ and $\phi_{h, (im)}$ from
	\begin{eqnarray*}
		p(\phi_{h, (re)} \mid \cdots) &\sim& Ga \left \{ a_2 +\frac{PQS}{2},  \sum_{q=h}^{Q} \psi_{q ,(re)}^{(h)} \sum_{p=1}^{P} \frac{\mb \alpha_{pq}'\mb \alpha_{pq}}{2 \tau_{pq, (re)}^2}\right \} ,\\
		p(\phi_{h, (im)} \mid \cdots) &\sim& Ga \left \{ a_2 +\frac{PQS}{2},  \sum_{q=h}^{Q} \psi_{q, (im)}^{(h)} \sum_{p=1}^{P} \frac{\mb \beta_{pq}'\mb \beta_{pq }}{2 \tau_{pq, (im)}^2}\right \},
	\end{eqnarray*}	
	where $\psi_{q,(\cdot)}^{(h)} = \prod_{t=1, t \neq h}^q \phi_{h, (\cdot)}$ for $h=1, \cdots, Q$ and $(\cdot)$  can be  $(re)~\text{or}~(im)$. 
	
	\item Sample the error variances $\sigma_{\epsilon,p}^2$, $p=1, \cdots, P$,  from the conditional posterior distribution
	\begin{equation*}
	p(\sigma_{\epsilon,p}^2 \mid \cdots) \sim IG \left \{   \frac{K+\nu}{2}, \sum_{k=1}^{K} | Y_{kp} - \Lambda_{k} \mb D_{k}|^2 + \frac{\nu}{g_{\epsilon,p}}   \right \},
	\end{equation*}
	and update the corresponding hyperparameters $g_{\epsilon,p}$
	\begin{equation*}
	p(g_{\epsilon,p} \mid \cdots) \sim IG \left \{   (\nu+1)/2, \nu/\sigma_{\epsilon,p}^2 + 1/G_{\epsilon}^2   \right \}.
	\end{equation*}				
\end{enumerate}

\section{Details of the SAMC Sampling Scheme}

We include additional subscripts on some previously defined parameters to make their dependence on the number of segments $L$ explicitly. For example, $\mb \xi_{L}=(\xi_{0,L}, \dots, \xi_{L,L})'$ denotes partition points. To provide more compact presentation, for $\ell=1, \cdots, L$,  we denote $PQS$-vectors of coefficients of basis functions as
\begin{eqnarray*}
	\mb \alpha_{\ell} &=& \{(\mb \alpha_{11 \ell}', \cdots, \mb \alpha_{P1 \ell}'), (\mb \alpha_{12 \ell}', \cdots, \mb \alpha_{P2 \ell}'), \cdots, (\mb \alpha_{1Q \ell}', \cdots, \mb \alpha_{PQ \ell}')\}',\\
	\mb \beta_{\ell} &=& \{(\mb \beta_{11 \ell}', \cdots, \mb \beta_{P1 \ell}'), (\mb \beta_{12 \ell}',\cdots, \mb \beta_{P2 \ell}'), \cdots, (\mb \beta_{1Q \ell}', \cdots, \mb \beta_{PQ \ell}')\}',
\end{eqnarray*}
and further let $\mb \alpha = (\mb \alpha_{1}', \cdots, \mb \alpha_{L}')'$ and $\mb \beta = (\mb \beta_{1}', \cdots, \mb \beta_{L}')'$.  Similarly,  we define $PQ$-vectors of smoothing parameters
\begin{eqnarray*}
	\mb \tau_{\ell,(re)}^2 &=& \{ [\tau_{11 \ell, (re)}^2, \cdots,  \tau_{P1 \ell, (re)}^2], [\tau_{12 \ell, (re)}^2, \cdots,  \tau_{P2 \ell, (re)}^2], \cdots, [\tau_{1Q \ell, (re)}^2, \cdots,  \tau_{PQ \ell, (re)}^2]\}',\\
	\mb \tau_{\ell,(im)}^2 &=& \{ [\tau_{11 \ell, (im)}^2, \cdots,  \tau_{P1 \ell, (im)}^2], [\tau_{12 \ell, (im)}^2, \cdots,  \tau_{P2 \ell, (im)}^2], \cdots, [\tau_{1Q \ell, (im)}^2, \cdots,  \tau_{PQ \ell, (im)}^2]\}',
\end{eqnarray*}
and $\mb \tau_{(re)}^2 = (\mb \tau_{1,(re)}^{2'}, \cdots, \mb \tau_{L,(re)}^{2'})'$ and $\mb \tau_{(im)}^2 = (\mb \tau_{1,(im)}^{2'}, \cdots, \mb \tau_{L,(im)}^{2'})'$. For the shrinkage parameters, we have
\begin{eqnarray*}
	\mb \phi_{\ell,(re)} &=& \{ \phi_{\ell 1, (re)}, \cdots,  \phi_{\ell Q, (re)}     \}',~ \mb \psi_{\ell,(re)} = \{ \psi_{\ell 1, (re)}, \cdots,  \psi_{\ell Q, (re)}     \}', \\
	\mb \phi_{\ell,(im)} &=& \{ \phi_{\ell  1, (im)}, \cdots,  \phi_{\ell Q, (im)}     \}',~ \mb \psi_{\ell,(im)} = \{ \psi_{\ell 1, (im)}, \cdots,  \psi_{\ell Q, (im)}     \}',
\end{eqnarray*}
and $\mb \phi_{(re)} = (\mb \phi_{1,(re)}', \cdots, \mb \phi_{L,(re)}')'$,  $\mb \phi_{(im)} = (\mb \phi_{1,(im)}', \cdots, \mb \phi_{L,(im)}')$, $\mb \psi_{(re)} = (\mb \psi_{1,(re)}', \cdots, \mb \psi_{L,(re)}')'$, and $\mb \psi_{(im)} = (\mb \psi_{1,(im)}', \cdots, \mb \psi_{L,(im)}')'$. In addition, we define $\mb D$ as the collection of all factors across segments and frequencies, such that $\mb D = \{ \mb D_{1}, \cdots, \mb D_{L}  \}$. Finally, we let the collection of parameters be $\Xi$, and denote current and newly drawn values by superscripts $c$ and $d$.

\subsection{Between-Model Moves}
For each iteration, we first propose a between-model move. Let the current value of the chain be $\left ( L^c,  \Xi^c_{L^c} \right)$. The chain is moved to proposed values $\left( L^d,  \Xi_{L^d}^d \right)$ by drawing from the proposal density $q (L^d,  \Xi_{L^d}^d \mid L^c, \Xi^c_{L^c})$. Specifically, the probability of accepting the draw is given by
\begin{equation*}
A = \min \left \{1,\frac{e^{\vartheta_{c,L^c}}}{e^{\vartheta_{{d},L^{d}}}} \frac{\pi(L^{d}, \Xi_{L^{d}}^{d} \mid \mb x) \times q (L^c, \Xi_{L^c}^c \mid L^{d},  \Xi^{d}_{L^{d}})}{\pi(L^c, \Xi_{L^c}^c \mid \mb x) \times q (L^{d}, \Xi_{L^{d}}^{d} \mid L^c, \Xi_{L^c}^c)}   \right \},
\end{equation*}
where  $\mb x$ is the time series observations and $\pi(\cdot)$ is the posterior distribution that is a product of the likelihood and prior distributions.  The specific form of the proposal density is
\begin{eqnarray*}
	q (L^d, \Xi_{L^d}^d \mid L^c, \Xi^c_{L^c})  &=&  q(L^d \mid L^c,  \Xi^c_{L^c}) \times q( \mb \xi_{L^d}^d \mid L^d, L^c, \Xi^c_{L^c}) \\
	&\times& q (\mb \tau_{(re),L^d}^{2d} \mid \mb \xi_{L^d}^d, L^d, L^c,  \Xi^c_{L^c} )
	\times q( \mb \tau_{(im),L^d}^{2d} \mid \mb \xi_{L^d}^d, L^d, L^c,  \Xi^c_{L^c} )\\
	&\times& q( \mb \phi_{(re),L^d}^d \mid \mb \xi_{L^d}^d, L^d, L^c,  \Xi^c_{L^c} )
	\times q( \mb \phi_{(im),L^d}^d \mid \mb \xi_{L^d}^d, L^d, L^c,  \Xi^c_{L^c} )\\
	&\times& q( \mb D_{L^d}^d \mid \mb \xi_{L^d}^d, L^d, L^c,  \Xi^c_{L^c}, \mb \tau_{(re),L^d}^{2d},  \mb \tau_{(im),L^d}^{2d}, \phi_{(im),L^d}^d  )\\
	&\times& q( \mb \alpha_{L^d}^d \mid \mb \xi_{L^d}^d, L^d, L^c,  \Xi^c_{L^c}, \mb \tau_{(re),L^d}^{2d},   \phi_{(re),L^d}^d, \mb D_{L^d}^d )\\
	&\times& q( \mb \beta_{L^d}^d \mid \mb \xi_{L^d}^d, L^d, L^c,  \Xi^c_{L^c},\mb \tau_{(im),L^d}^{2d},   \phi_{(im),L^d}^d, \mb D_{L^d}^d ),
\end{eqnarray*}
which means, given $L^c$ and $\Xi_{L^c}^c$, the parameters $L^d$, $\mb \tau_{(re),L^d}^{2d}$, $\mb \tau_{(im),L^d}^{2d}$,  $\mb \phi_{(re),L^d}^d$, $\mb \phi_{(im),L^d}^d$, $\mb D$, $\mb \alpha_{L^d}^d$, and $\mb \beta_{L^d}^d$ are sequentially drawn. Details on each of these quantities are described below.

The first step in the between-model move is proposing the number of partitions (i.e. choosing between a birth or a death move). Let $L_{\max}$ be the a priori maximum number of segments, and $L_{2\min}^c$ be the current number of segments which have at least $2t_{\min}$ observations. The number of segments $L^d=k$ is drawn from
\[  q(L^d = k \mid L^c,  \Xi^c_{L^c}) = \begin{cases}
1/2 & \text{if}~ k=L^c-1~ \text{or}~ L^c+1, L^c \ne 1, L^c \ne L_{\max}, L^c_{2\min} \ne 0, \\
1 &  \text{if}~ k=L^c-1, L^c=L_{\max} ~\text{or}~ L^c_{2\min} = 0, \\
1 & \text{if}~ k=L^c+1, L^c=1.
\end{cases}
\]

The second step involves drawing $\mb \tau_{(re),L^d}^{2d}$, $\mb \tau_{(im),L^d}^{2d}$,  $\mb \phi_{(re),L^d}^d$, $\mb \phi_{(im),L^d}^d$, $\mb D$, $\mb \alpha_{L^d}^d$, and $\mb \beta_{L^d}^d$ in order, depending on whether the proposed move is a birth move or a death move.

{\bf Birth:} A birth move is proposed if $L^d = L^c + 1$.  Then, a new partition is proposed as $$\mb \xi_{L^d}^d =( \xi_{0,L^c}^c, \dots, \xi_{z^*-1,L^c}^c, \xi_{z^*,L^d}^d, \xi_{z^*,L^c}^c, \dots, \xi_{L^c,L^c}^c).$$ In particular,
\begin{eqnarray*}
	q( \xi_{q,L^d}^d=t^* \mid L^d, L^c, \mb \xi_{L^c}^c) &=& \pr(q=z^*\mid L^d, L^c, \mb \xi_{L^c}^c) \\
	&\times& \pr(\xi_{z^*,L^d}^d =t^* \mid q=z^*, L^d, L^c, \mb \xi_{L^c}^c)\\
	~ &=& (1/L^c_{2\min}) \times \left \{  1/(t_{z^*, L^c}- 2t_{\min} + 1)   \right\},
\end{eqnarray*}
which means that the new partition point is proposed by first sampling a current segment $q=z^*$ to be split, and then randomly selecting the new partition point $t^*$ within the segment, subject to the constraint $\xi_{z^*-1,L^c}^c + t_{\min} < t^* < \xi_{z^*,L^c}^c - t_{\min}$.

Next, given the newly proposed partition point $\xi^d_{z^*,L^d}$, we propose the smoothing parameters for the real and imagary parts
\begin{eqnarray*}
	\mb \tau^{2d}_{(re), L^d} &=&(\mb \tau^{2'c}_{1,L^c, (re)}, \dots \mb \tau^{2'c}_{z^*-1,L^c,(re)}, \mb \tau^{2'd}_{z^*,L^d,(re)}, \mb \tau^{2'd}_{z^*+1,L^d,(re)}, \mb \tau^{2'c}_{z^*+1,L^c, (re)}, \dots \mb \tau^{2'c}_{L^c, L^c, (re)})',\\
	\mb \tau^{2d}_{(im), L^d} &=&(\mb \tau^{2'c}_{1,L^c, (im)}, \dots \mb \tau^{2'c}_{z^*-1,L^c,(im)}, \mb \tau^{2'd}_{z^*,L^d,(im)}, \mb \tau^{2'd}_{z^*+1,L^d,(im)}, \mb \tau^{2'c}_{z^*+1,L^c, (im)}, \dots \mb \tau^{2'c}_{L^c, L^c, (im)})'.
\end{eqnarray*}
Let $\mb U_{\tau_{(re)}^2}$ and $\mb U_{\tau_{(im)}^2}$ be random $PQ$-vectors whose elements are $u_j/(1-u_j)$ for $j=1, \cdots, PQ$, where $u_j$ are drawn independently from Uniform$(a,1-a)$.  Then, the smoothing parameters of the real and imaginary parts of the basis function coefficients in the newly proposed two segments are drawn as
\begin{eqnarray*}
	\mb \tau_{z^*,L^d, (re)}^{2'd} = \mb U_{\tau_{(re)}^{2}}	\circ \mb \tau_{z^*,L^c,(re)}^{2'c}, && \mb \tau_{z^*+1,L^d,(re)}^{2'd} = (\mb U_{\tau_{(re)}^{2}})^{-1}	\circ \mb \tau_{z^*,L^c}^{2'c},\\
	\mb \tau_{z^*,L^d, (im)}^{2'd} = \mb U_{\tau_{(im)}^{2}}	\circ \mb \tau_{z^*,L^c,(im)}^{2'c}, && \mb \tau_{z^*+1,L^d, (im)}^{2'd} = (\mb U_{\tau_{(im)}^{2}})^{-1}	\circ \mb \tau_{z^*,L^c}^{2'c},
\end{eqnarray*}	
where $\circ$ represents the Schur or elementwise product of vectors \citep{green1995}. In a similar fashion, we propose the shrinkage parameters
\begin{eqnarray*}
	\mb \phi^{d}_{(re), L^d} &=&(\mb \phi^{'c}_{1,L^c, (re)}, \dots \mb \phi^{'c}_{z^*-1,L^c,(re)}, \mb \phi^{'d}_{z^*,L^d,(re)}, \mb \phi^{2'd}_{z^*+1,L^d,(re)}, \mb \phi^{'c}_{z^*+1,L^c, (re)}, \dots \mb \phi^{'c}_{L^c, L^c, (re)})',\\
	\mb \phi^{d}_{(im), L^d} &=&(\mb \phi^{'c}_{1,L^c, (im)}, \dots \mb \phi^{'c}_{z^*-1,L^c,(im)}, \mb \phi^{'d}_{z^*,L^d,(im)}, \mb \phi^{2'd}_{z^*+1,L^d,(im)}, \mb \phi^{'c}_{z^*+1,L^c, (im)}, \dots \mb \phi^{'c}_{L^c, L^c, (im)})',
\end{eqnarray*}
and
\begin{eqnarray*}
	\mb \phi_{z^*,L^d, (re)}^{'d} = \mb U_{\phi_{(re)}}	\circ \mb \phi_{z^*,L^c,(re)}^{'c}, && \mb \phi_{z^*+1,L^d,(re)}^{'d} = (\mb U_{\phi_{(re)}})^{-1}	\circ \mb \phi_{z^*,L^c}^{'c},\\
	\mb \phi_{z^*,L^d, (im)}^{'d} = \mb U_{\phi_{(im)}}	\circ \mb \phi_{z^*,L^c,(im)}^{'c}, && \mb \phi_{z^*+1,L^d,(im)}^{'d} = (\mb U_{\phi_{(im)}})^{-1}	\circ \mb \phi_{z^*,L^c}^{'c},
\end{eqnarray*}	
where $\mb U_{\phi_{(re)}}$ and $\mb U_{\phi_{(im)}}$ are random $Q$-vectors whose elements are $u_j/(1-u_j)$ for $j=1, \cdots, Q$, and the $u_j$ are drawn independently from Uniform$(a,1-a)$.

The factors
\begin{equation*}
\mb D_{L^d}^{'d} = (\mb D_{1,L^d}^{'d}, \dots, \mb D_{z^*-1,L^c}^{'c}, \mb D_{z^*,L^d}^{'d}, \mb D_{z^*+1,L^d}^{'d}, \mb D_{z^*+1,L^c}^{'c}, \dots, \mb \alpha_{L^c,L^c}^{'c})'
\end{equation*}
is proposed from $q( \mb D_{L^d}^d \mid \mb \xi_{L^d}^d, L^d, L^c,  \Xi^c_{L^c}, \mb \tau_{(re),L^d}^{2'd},  \mb \tau_{(im),L^d}^{2'd}, \phi_{(im),L^d}^d)$. In particular, similar to \eqref{D}, $ \mb D_{z^*,L^d}^{'d}$ and $\mb D_{z^*+1,L^d}^{'d}$ are drawn from
\begin{eqnarray}
\label{D_1}
p(\mb D_{kz^*,L^d}^{'d} \mid \cdots) &\sim& \text{CN} \left (\mb \mu_{D_{kz^*,L^d}^{d}}, \Sigma_{D_{kz^*,L^d}^{d}} \right ),  k=1, \cdots, K_{z^*},\\
\label{D_2}
p(\mb D_{kz^*+1,L^d}^{'d} \mid \cdots) &\sim& \text{CN} \left (\mb \mu_{D_{kz^*+1,L^d}^{d}}, \Sigma_{D_{kz^*+1,L^d}^{d}} \right ),  k=1, \cdots, K_{z^*+1}
\end{eqnarray}
where $$\Sigma_{D_{kz^*,L^d}^{d}} = (\Lambda_{k z^*, L^d}^{*d} \Sigma_{\epsilon}^{-1} \Lambda_{k z^*, L^d}^{d} + I_{Q})^{-1},~ \mb \mu_{D_{kz^*,L^d}^{d}} = \Sigma_{D_{kz^*,L^d}^{d}} \Lambda_{k z^*, L^d}^{*d} \Sigma_{\epsilon}^{-1} \mb Y_{k z^*,L},$$
$$\Sigma_{D_{kz^*+1,L^d}^{d}} = (\Lambda_{k z^*+1, L^d}^{*d} \Sigma_{\epsilon}^{-1} \Lambda_{k z^*+1, L^d}^{d} + I_{Q})^{-1},~ \mb \mu_{D_{kz^*+1,L^d}^{d}} = \Sigma_{D_{kz^*+1,L^d}^{d}} \Lambda_{k z^*+1, L^d}^{*d} \Sigma_{\epsilon}^{-1} \mb Y_{k z^*+1, L}.$$

Then, vectors of the real parts of basis functions coefficients
\begin{eqnarray*}
	\mb \alpha_{L^d}^{'d} &=& (\mb \alpha_{1,L^d}^{'d}, \dots, \mb \alpha_{z^*-1,L^c}^{'c}, \mb \alpha_{z^*,L^d}^{'d}, \mb \alpha_{z^*+1,L^d}^{'d}, \mb \alpha_{z^*+1,L^c}^{'c}, \dots, \mb \alpha_{L^c,L^c}^{'c})',
\end{eqnarray*}
is proposed from $q( \mb \alpha_{L^d}^d \mid \mb \xi_{L^d}^d, L^d, L^c,  \Xi^c_{L^c}, \mb \tau_{(re),L^d}^{2p},   \phi_{(re),L^d}^d, \mb D_{L^d}^d )$. The $pq$th elements of $\mb \alpha_{z^*,L^d}^{'d}$ and $\mb \alpha_{z^*+1,L^d}^{'d}$ are drawn from the conditional normal distribution shown in \eqref{RE}, for $p=1, \cdots, P$ and $q=1, \cdots, Q$,
\begin{eqnarray}
\label{alpha_1}
p(\mb \alpha_{pq z^*, L^d} \mid \cdots) &\sim& \text{N} (\mb \mu_{\alpha_{pq z^*, L^d}}, \Sigma_{\alpha_{pq z^*, L^d}}), \\
\label{alpha_2}
p(\mb \alpha_{pq z^*+1, L^d} \mid \cdots) &\sim& \text{N} (\mb \mu_{\alpha_{pq z^*+1, L^d}}, \Sigma_{\alpha_{pq z^*+1, L^d}}),
\end{eqnarray}
where,
$$\Sigma_{\alpha_{pq z^*, L^d}} = \left [  \Omega_{p q z^*, L^d}^{(re)^{-1}} + 2\sigma_{\epsilon,p}^{-2} \sum_{k=1}^K |D_{k z^*, L^d}^{(q)}|^2 \mb \omega_k^{(re)} \mb \omega_k^{(re)'} \right ]^{-1},$$
$$\mb \mu_{\alpha_{pq z^*, L^d}} = \frac{2}{\sigma_{\epsilon,d}^2} \Sigma_{\alpha_{pq z^*, L^d}} \left [  \sum_{k=1}^K \Re\{ - Y_{k p z^*, L^d}^* D_{k z^*, L^d}^{(q)} + S_{k z^*, L^d}^{(d+\backslash q)} D_{k z^*, L^d}^{(q)*}  \} \mb \omega_k^{(re)}    \right ], $$
$$\Sigma_{\alpha_{pq z^*+1, L^d}} = \left [  \Omega_{p q z^*+1, L^d}^{(re)^{-1}} + 2\sigma_{\epsilon,p}^{-2} \sum_{k=1}^K |D_{k z^*+1, L^d}^{(q)}|^2 \mb \omega_k^{(re)} \mb \omega_k^{(re)'} \right ]^{-1},$$
$$\mb \mu_{\alpha_{pq z^*+1, L^d}} = \frac{2}{\sigma_{\epsilon,d}^2} \Sigma_{\alpha_{pq z^*+1, L^d}} \left [  \sum_{k=1}^K \Re\{ - Y_{k p z^*+1, L^d}^* D_{k z^*+1, L^d}^{(q)} + S_{k z^*+1, L^d}^{(p+\backslash q)} D_{k z^*+1, L^d}^{(q)*}  \} \mb \omega_k^{(re)}    \right ]. $$
Vectors of the imaginary parts of basis functions coefficients
\begin{eqnarray*}
	\mb \beta_{L^d}^{'d} &=& (\mb \beta_{1,L^d}^{'d}, \dots, \mb \beta_{z^*-1,L^c}^{'c}, \mb \beta_{z^*,L^d}^{'d}, \mb \beta_{z^*+1,L^d}^{'d}, \mb \beta_{z^*+1,L^c}^{'c}, \dots, \mb \beta_{L^c,L^c}^{'c})',
\end{eqnarray*}
can be drawn similarly, and thus omitted.

Lastly, the proposed move is accepted with probability $A=\min \left \{ 1, \Pi  \right \}$, where
\begin{eqnarray*}
	\Pi &=& \frac{e^{\vartheta_{c,L^c}}}{e^{\vartheta_{{d},L^{d}}}}  \frac{p(\Xi^d_{L^d} \mid \mb x_t, L^d) p (\Xi_{L^d}^d \mid L^d) p(L^d)}{p (\Xi^c_{L^c} \mid \mb x_t, L^c) p (\Xi_{L^c}^c \mid L^c) p(L^c)} \\
	&\times& \frac{p(L^c \mid L^d)   } {p(L^d \mid L^c) p (\mb \xi^d_{L^d}\mid L^d, L^c)}
	\times   \frac{p (\mb D^{c}_{z^*,L^c})}{ p (\mb D^{d}_{z^*,L^d}) p (\mb D^{d}_{z^*+1,L^d})}
	\times    \frac{p (\mb \alpha^{c}_{z^*,L^c})}{ p (\mb \alpha^{d}_{z^*,L^d}) p (\mb \alpha^{d}_{z^*+1,L^d})}
	\times    \frac{p (\mb \beta^{c}_{z^*,L^c})}{ p (\mb \beta^{d}_{z^*,L^d}) p (\mb \beta^{d}_{z^*+1,L^d}) }  \\
	&\times& \frac{ |\mb J|}{p(\mb U_{\tau^{2}_{(re)}}) p(\mb U_{\tau^2_{(im)}}) p(\mb U_{\phi_{(re)}}) p(\mb U_{\phi_{(im)}})},
\end{eqnarray*}
where $p(\mb U_{(\cdot)})$  is the density of uniform distribution $[a, 1-a]$, and $p(\mb D^{d}_{z^*,L^d})$ and $p(\mb D^{d}_{z^*+1,L^d})$, $p(\mb \alpha^{d}_{z^*,L^d})$, $p(\mb \alpha^{d}_{z^*+1,L^d})$, $p(\mb \beta^{d}_{z^*,L^d})$ and $p(\mb \beta^{d}_{z^*+1,L^d})$ are the Normal densities presented in \eqref{D_1},\eqref{D_2}, \eqref{alpha_1}, and \eqref{alpha_2}, respectively.  $|\mb J|$ is the Jacobian such that
\begin{eqnarray*}
	|\mb J| &=& 2 ( \mb \tau_{z^*,L^d,(re)}^{2d} +     \mb \tau_{z^*+1,L^d,(re)}^{2d}    )'  ( \mb \tau_{z^*,L^d,(re)}^{2d} +     \mb \tau_{z^*+1,L^d,(re)}^{2d}    ) \\
	&+& 2 ( \mb \tau_{z^*,L^d,(im)}^{2d} +     \mb \tau_{z^*+1,L^d,(im)}^{2d}    )'  ( \mb \tau_{z^*,L^d,(im)}^{2d} +     \mb \tau_{z^*+1,L^d,(im)}^{2d}    ) \\
	&+& 2 ( \mb \phi_{z^*,L^d,(re)}^{d} +     \mb \phi_{z^*+1,L^d,(re)}^{d}    )'  ( \mb \phi_{z^*,L^d,(re)}^{d} +     \mb \phi_{z^*+1,L^d,(re)}^{d}    ) \\
	&+& 2 ( \mb \phi_{z^*,L^d,(im)}^{d} +     \mb \phi_{z^*+1,L^d,(im)}^{d}    )'  ( \mb \phi_{z^*,L^d,(im)}^{d} +     \mb \phi_{z^*+1,L^d,(im)}^{d}    ).
\end{eqnarray*}

{\bf Death}:  A death move is proposed if $L^d=L^c-1$,  a vector of partitions $$\mb \xi_{L^p}^p =(\xi_{0,L^c}^c, \dots, \xi_{z^*-1,L^c}^c,  \xi_{z^*+1,L^c}^c, \dots, \xi_{L^c,L^c}^c)$$ is sampled by drawing one of the partition to remove with equal probability so that the probability of the $z^*$th partition being sampled to removed is  $q(\xi_{z^*,L^d}^d \mid L^d, L^c, \mb \xi_{L^c}^c) = 1/(L^c-1)$.  The smoothing parameters for the combined segment are proposed as  $\mb \tau_{z^*, L^d, (re)}^{2d} = \sqrt{ \mb \tau_{z^*,L^c, (re)}^{2c} \circ \mb \tau_{z^*+1,L^c, (re)}^{2c} }$ and $\mb \tau_{z^*, L^d, (im)}^{2d} = \sqrt{ \mb \tau_{z^*,L^c, (im)}^{2c} \circ \mb \tau_{z^*+1,L^c, (im)}^{2c} }$  so that the proposed vectors of all smoothing parameters are
\begin{eqnarray*}
	\mb \tau^{2d}_{L^d, (re)} =(\mb \tau^{2'c}_{1,L^c, (re)}, \dots \mb \tau^{2'c}_{z^*-1,L^c, (re)}, \mb \tau^{2'd}_{z^*,L^d, (re)},  \mb \tau^{2'c}_{z^*+2,L^c, (re)}, \dots \mb \tau^{2'c}_{L^c,L^c, (re)}),\\
	\mb \tau^{2d}_{L^d, (im)} =(\mb \tau^{2'c}_{1,L^c, (im)}, \dots \mb \tau^{2'c}_{z^*-1,L^c, (im)}, \mb \tau^{2'd}_{z^*,L^d, (im)},  \mb \tau^{2'c}_{z^*+2,L^c, (im)}, \dots \mb \tau^{2'c}_{L^c,L^c, (im)}).
\end{eqnarray*}
Similarly, the shrinkage parameters for the combined segment are taken as $\mb \phi_{z^*, L^d, (re)}^{d} = \sqrt{ \mb \phi_{z^*,L^c, (re)}^{c} \circ \mb \phi_{z^*+1,L^c, (re)}^{c} }$ and $\mb \phi_{z^*, L^d, (im)}^{d} = \sqrt{ \mb \phi_{z^*,L^c, (im)}^{c} \circ \mb \phi_{z^*+1,L^c, (im)}^{c} }$. Thus, the proposed vectors of all shrinkage parameters are
\begin{eqnarray*}
	\mb \phi^{d}_{L^d, (re)} =(\mb \phi^{'c}_{1,L^c, (re)}, \dots \mb \phi^{'c}_{z^*-1,L^c, (re)}, \mb \phi^{'d}_{z^*,L^d, (re)},  \mb \phi^{'c}_{z^*+2,L^c, (re)}, \dots \mb \phi^{'c}_{L^c,L^c, (re)}),\\
	\mb \phi^{p}_{L^d, (im)} =(\mb \phi^{'c}_{1,L^c, (im)}, \dots \mb \phi^{'c}_{z^*-1,L^c, (im)}, \mb \phi^{'d}_{z^*,L^d, (im)},  \mb \phi^{'c}_{z^*+2,L^c, (im)}, \dots \mb \phi^{'c}_{L^c,L^c, (im)}).
\end{eqnarray*}

Then, vector of basis function coefficients
\begin{eqnarray*}
	\mb \alpha_{L^d}^{d} &=& (\mb \alpha_{1,L^d}^{'d}, \dots, \mb \alpha_{z^*-1,L^c}^{'c}, \mb \alpha_{z^*,L^d}^{'d},  \mb \alpha_{z^*+2,L^c}^{'c}, \dots, \mb \alpha_{L^c,L^c}^{'c})', \\
	\mb \beta_{L^d}^{d} &=& (\mb \beta_{1,L^d}^{'d}, \dots, \mb \beta_{z^*-1,L^c}^{'c}, \mb \beta_{z^*,L^d}^{'d},  \mb \beta_{z^*+2,L^c}^{'c}, \dots, \mb \beta_{L^c,L^c}^{'c})',
\end{eqnarray*}
are proposed from $$q( \mb \alpha_{L^d}^d \mid \mb \xi_{L^d}^d, L^d, L^c,  \Xi^c_{L^c}, \mb \tau_{(re),L^d}^{2d},   \mb \phi_{(re),L^d}^d, \mb D_{L^d}^d ),$$
$$q( \mb \beta_{L^d}^d \mid \mb \xi_{L^d}^d, L^d, L^c,  \Xi^c_{L^c},\mb \tau_{(im),L^d}^{2d},   \mb \phi_{(im),L^d}^d, \mb D_{L^d}^d ),$$ respectively.  In particular, the $pq$th element of $\mb \alpha_{z^*,L^d}^{d}$ and $\mb \beta_{z^*,L^d}^{d}$ are drawn from the Normal desnities described in the birth step. The acceptance probability is $A = \min \{ 1, 1/ \Pi \}$, where $\Pi$ is the same in the birth step.

\subsection{Within-model moves}
The within-model moves involve no change in the number of partitions or segments, i.e. $L^d=L^c$. For fixed $L$, we first relocate a partition point, and then, conditional on the new location of partition, factors and basis functions coefficients involved are updated. We either accept or reject these two steps jointly.

We first select a partition, $\xi_{z^*, L}$, from $L-1$ possible partitions, and then relocate it to another time point, where the position has to be in the interval $[\xi_{z^*-1,L}, \delta_{z^*+1,L}]$ and satisfy the condition that the new location is at least $t_{\min}$ away from $\xi_{z^*-1, L}$ and $\xi_{z^*+1, L}$. Thus, we draw new partition from
\begin{eqnarray*}
	q(\xi^d_{z^*, L}=t) &=& \pr(q=z^*) \pr(\delta^d_{z^*, L}=t \mid q=z^*) \\
	&=& 1/(L^c-1) \pr(\xi^d_{z^*, L}=t \mid q=z^*),
\end{eqnarray*}
and use a mixture distribution for $\pr(\xi^d_{z^*, L}=t \mid q=z^*)$ \citep{rosen2012} so that the sampler can explore parameter space efficiently, such that
\begin{equation*}
\pr(\xi^d_{z^*, L}=t \mid q=z^*) =p \times \pr_1(\xi^d_{z^*, L}=t \mid \xi^c_{z^*, L}) + (1-p) \times \pr_2(\xi^c_{z^*, L}=t \mid \xi^c_{z^*, L}),
\end{equation*}
where $p \in (0,1)$ is a fixed parameter. we set $p=0.2$ for most of our numerical examples to balance of high acceptance rate and ability of exploring parameter spaces. The first probability $\pr_1$ corresponds to the big partition relocation so that the algorithm can explore whole parameter space, such that $\ \pr_1(\xi^d_{z^*,L}=t \mid \xi^c_{z^*,L})=(t_{z^*}+t_{z^*+1}-2t_{\min})^{-1}$ and  is subject to $\xi_{z^*-1,L} + t_{\min} < t < \xi_{z^*,L} - t_{\min}$.  The second probability $\pr_2$ is associated with the small partition relocation such that we only allow the partition stays or relocate to its neighbor with equal probability.

We then update $\mb D_{z^*, L}^d$ and $\mb D_{z^*+1, L}^d$ from the Normal distribution as in \eqref{D_1} and \eqref{D_2}, respectively. Conditional on newly drawn factors, in a similar fashion to \eqref{alpha_1} and \eqref{alpha_2}, $\mb \alpha_{*}^d = (\mb \alpha_{z^*, L}^{'d}, \mb \alpha_{z^*+1, L}^{'d})'$ and $\mb \beta_{*}^d = (\mb \beta_{z^*, L}^{'d}, \mb \beta_{z^*+1, L}^{'d})'$, are independently drawn. Denote $\mb \tau_{*}^{2d} = (\mb \tau_{z^*, L,(re)}^{2'd}, \mb \tau_{z^*+1, L, (re)}^{2'd}, \mb \tau_{z^*, L,(im)}^{2'd}, \mb \tau_{z^*+1, L, (im)}^{2'd})'$ as collection of all smoothing parameter involved, and $\mb \phi_{*}^{d} = (\mb \phi_{z^*, L,(re)}^{'d}, \mb \phi_{z^*+1, L, (re)}^{'d}, \mb \phi_{z^*, L,(im)}^{'d}, \mb \phi_{z^*+1, L, (im)}^{'d})'$ as collection of all shrinkage parameter involved.  The new draw is accepted with probability
\begin{equation*}
A = \min \left \{ 1, \frac{\pi(\mb x^d_{*}  \mid \mb \alpha_*^d, \mb \beta_*^d, \mb D_{z^*, L}^d, \mb D_{z^*, L+1}^d )  q( \mb \alpha_*^c, \mb \beta_*^c, \mb D_{z^*, L}^c, \mb D_{z^*, L+1}^c \mid \mb x_*^c, \mb \tau_{*}^{2c},  \mb \phi_{*}^{c})}{\pi(\mb x^c_{*}  \mid \mb \alpha_*^c, \mb \beta_*^c, \mb D_{z^*, L}^c, \mb D_{z^*, L+1}^c ) q(\mb \alpha_*^d, \mb \beta_*^d, \mb D_{z^*, L}^d, \mb D_{z^*, L+1}^d \mid \mb x_*^p, \mb \tau_{*}^{2d},  \mb \phi_{*}^{d})}   \right \}.
\end{equation*}
where $\pi(\cdot)$ is a target distribution that is a product of likelihood function and prior distributions, and $q(\cdot)$ is proposal distritution.

Next, the smoothing parameters of all segments are updated through Gibbs sampling procedures described in \eqref{tau1} and \eqref{tau2}, and the corresponding hyperparameters are drawn as in \eqref{g1} and \eqref{g2}. Lastly, sample error variance $\sigma_{\epsilon,p}^2$, $p=1, \cdots, P$,  from conditional posterior distribution
\begin{equation*}
p(\sigma_{\epsilon,p}^2 \mid \cdots) \sim \text{IG} \left \{  \sum_{\ell=1}^L \frac{K+\nu}{2}, \sum_{\ell=1}^{L} \sum_{k=1}^{K} | Y_{k \ell p} - \Lambda_{k \ell} \mb D_{k \ell}|^2 + \frac{\nu}{g_{\epsilon,p}}   \right \},
\end{equation*}
and  the corresponding hyperparameters $g_{\epsilon,p}$ are updated from
\begin{equation*}
p(g_{\epsilon,p} \mid \cdots) \sim \text{IG} \left \{   (\nu+1)/2, \nu/\sigma_{\epsilon,p}^2 + 1/G_{\epsilon}^2   \right \}.
\end{equation*}

\end{document}